\documentclass[12pt,preprint]{aastex}
\begin{document}

\title{Tracing PAHs and Warm Dust Emission in the Seyfert Galaxy
NGC~1068}
\author{
Justin H. Howell,\altaffilmark{1}
Joseph M. Mazzarella,\altaffilmark{1}
Ben H.P. Chan,\altaffilmark{1}
Steven Lord,\altaffilmark{2}
Jason A. Surace,\altaffilmark{3}
David T. Frayer,\altaffilmark{2}
P~.N.~Appleton,\altaffilmark{2}
Lee Armus,\altaffilmark{3}
Aaron S. Evans,\altaffilmark{4}
Greg Bothun,\altaffilmark{5}
Catherine M. Ishida,\altaffilmark{6}
Dong-Chan Kim,\altaffilmark{7}
Joseph B. Jensen,\altaffilmark{8}
Barry F. Madore\altaffilmark{1,9}
David B. Sanders,\altaffilmark{10}
Bernhard Schulz,\altaffilmark{2}
Tatjana Vavilkin,\altaffilmark{4}
Sylvain Veilleux,\altaffilmark{7}
Kevin Xu\altaffilmark{2}
}
\altaffiltext{1}{Infrared Processing \& Analysis Center, MS 100-22, 
California Institute of Technology,
Pasadena, CA 91125; jhhowell,mazz,bchan@ipac.caltech.edu}
\altaffiltext{2}{NASA Herschel Science Center, IPAC, MS 100-22, 
California Institute of Technology,
Pasadena, CA 91125; lord,frayer,apple,bschulz,cxu@ipac.caltech.edu}
\altaffiltext{3}{Spitzer Science Center, MS 314-6, California Institute 
of Technology, Pasadena,
CA 91125; jason,lee@ipac.caltech.edu}
\altaffiltext{4}{Department of Physics and Astronomy,
State University of New York at Stony Brook, Stony Brook, NY 11794-3800;
\ aevans@orchid.ess.sunysb.edu, vavilkin@grad.physics.sunysb.edu}
\altaffiltext{5}{University of Oregon, Physics Department, Eugene OR, 97402;
nuts@bigmoo.uoregon.edu}
\altaffiltext{6}{Subaru Telescope, National Astronomical Observatory of 
Japan, Hilo, HI 96720; cat@subaru.naoj.org}
\altaffiltext{7}{Department of Astronomy, University of Maryland, 
College Park, MD 20742; kim,veilleux@astro.umd.edu}
\altaffiltext{8}{Gemini Observatory, 950 North Cherry Avenue, Tucson, AZ 
85719; jjensen@gemini.edu}
\altaffiltext{9}{The Observatories, Carnegie Institution of Washington,
813 Santa Barbara Street, Pasadena, CA 91101; barry@ociw.edu}
\altaffiltext{10}{Institute for Astronomy, University of Hawaii, 2680
Woodlawn Drive, Honolulu, HI 96822; sanders@ifa.hawaii.edu}

\begin{abstract}

We present a study of the nearby Seyfert galaxy NGC 1068 using mid- and far-
infrared data acquired with the IRAC, IRS, and MIPS instruments aboard the
Spitzer Space Telescope. The images show extensive $8~\mu$m and $24~\mu$m 
emission coinciding with star formation in the inner spiral 
approximately 15\arcsec (1 kpc) 
from the nucleus, and a bright complex of star formation $\sim$47\arcsec\ 
(3 kpc) SW of the nucleus.  The brightest $8~\mu$m PAH emission regions 
coincide remarkably well with knots observed in an H$\alpha$ image.  Strong 
PAH features at 6.2, 7.7, 8.6, and $11.3~\mu$m are detected in IRS spectra 
measured at numerous locations inside, within, and outside the inner spiral.  
The IRAC colors and IRS spectra of these regions rule out dust heated by the 
AGN as the primary emission source; the SEDs are dominated by starlight and PAH
emission.  The equivalent widths and flux ratios of the PAH features in the
inner spiral are generally consistent with conditions in a typical spiral galaxy
ISM.  Interior to the inner spiral, the influence of the AGN on the ISM is 
evident via PAH flux ratios indicative of a higher ionization parameter and a
significantly smaller mean equivalent width than observed in the inner spiral.
The brightest 8 and $24~\mu$m emission peaks in the disk of the galaxy,
even at distances beyond the inner spiral, are located within
the ionization cones traced by [O III]/H$\beta$, and they are also
remarkably well aligned with the axis of the radio jets. Although it is
possible that radiation from the AGN may directly enhance PAH
excitation or trigger the formation of OB stars that subsequently excite PAH
emission at these locations in the inner spiral, the orientation of collimated
radiation from the AGN and star formation knots in the inner spiral could be
coincidental. The brightest PAH and $24~\mu$m emitting regions 
are also located precisely where two spiral arms of molecular
gas emerge from the ends of the inner stellar bar; this is consistent with
kinematic models that predict maxima in the accumulation and compression of
the ISM where gas gets trapped within the inner Lindblad resonance of a
large stellar bar that contains a smaller, weaker bar.
\end{abstract}


\section{Introduction}

Polycyclic aromatic hydrocarbons (PAHs) appear to be ubiquitous in the 
ISM of massive disk galaxies.  PAHs emit copiously near regions of active star 
formation where molecules are heated by ultraviolet radiation and are 
usually  associated with the photo-dissociation regions in molecular clouds.  
The infrared spectra of active galaxies generally lack the PAH emission lines 
characteristic of starburst galaxies \citep{roche91}. 
\citet*{freudling03} reported the detection of hot dust and weak PAHs 
in type 1 AGN, and cooler dust and strong PAHs in type 2 AGN, similar to the
results of \citet{clavel}, \citet{buchanan06}, and \citet{apple07}.  
These results are interpreted as an indication 
that the high energy photons from the AGN destroy PAH molecules exposed to 
extreme UV radiation from 
the nucleus, with any observed PAH emission coming from star forming regions 
shielded by intervening material \citep{voit92, allain96, maloney99, 
freudling03}.
The {\it Spitzer} Space Telescope provides the opportunity to investigate 
the distribution of PAH emission in a variety of galaxies through its 
correlation with the zone of influence of the AGN and/or star-forming regions.

\begin{figure}
\vbox{
\begin{center}
\mbox{\includegraphics[width=3.5in]{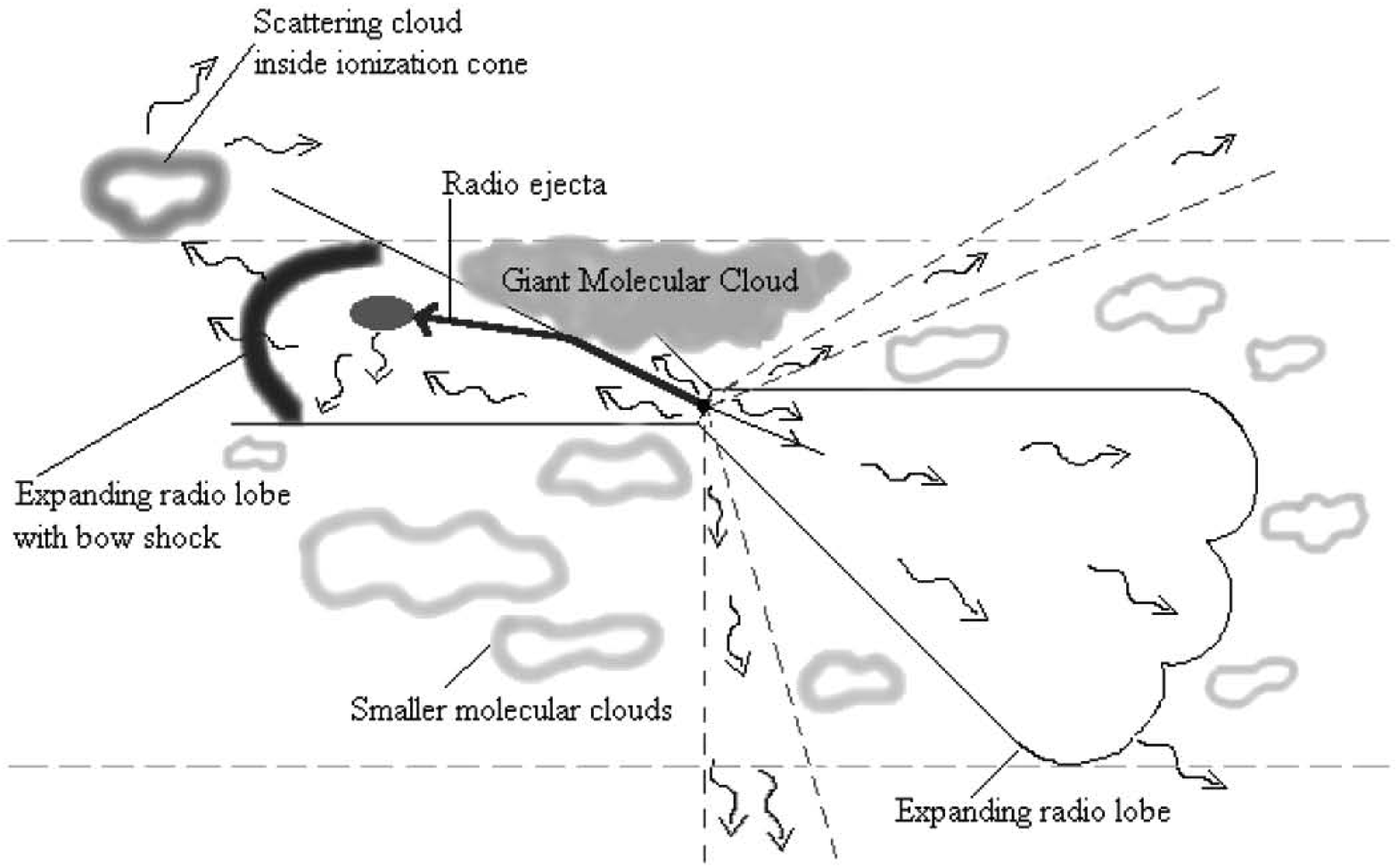}\quad
        \includegraphics[width=3.5in]{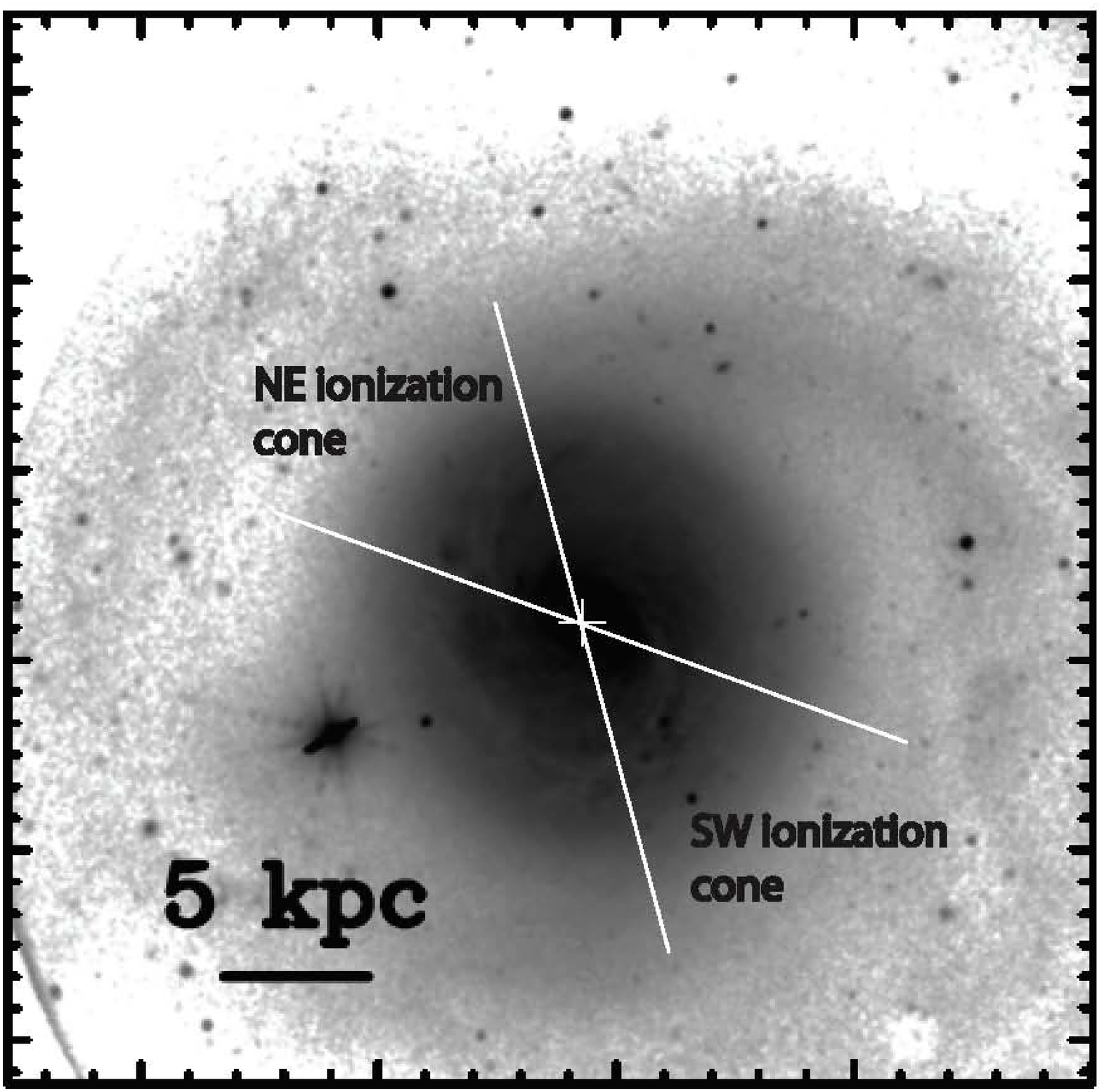}}
\figcaption{\small
Diagram of NGC~1068 ionization cone.  Left panel: view in the plane 
of the disk, from \citet{bruhweiler01}.  Right panel: R0 continuum 
image with ionization cones illustrated (adapted from \citet{veilleux03}).
Large tick marks are at $1^{\prime}$ intervals.
\label{cartoon}
}
\end{center}}
\end{figure}

NGC~1068 is a prototypical nearby Seyfert 2 galaxy \citep{antonucci85}, 
and was the first AGN to be detected \citep{fath09}.
With an infrared luminosity $\rm L_{ir} = 10^{11.3}~L_{\odot}$, NGC~1068 
is a luminous infrared galaxy (LIRG) and is a member of the IRAS
Revised Bright Galaxy Sample (RBGS; Sanders et al. 2003) consisting of
extragalactic objects with $60\mu$m flux greater than 5.24~Jy.
In addition to the AGN, the galaxy contains a bright inner spiral with 
active star formation approximately 1~kpc from the nucleus; this structure 
has often been referred to as a ``starburst ring" \citep{scoville, schinnerer, 
spinoglio05}.  \citet{td88} estimated that star formation would consume the 
available gas within 0.5~Gyr at current star formation rates.
\citet{yuankuo98} showed that the inner spiral resides at the 
outer Lindblad resonance.  The inner spiral arms begin at the ends of a
stellar bar \citep{schinnerer}, at the location of peak collisional 
energy dissipation \citep{combes85}.
Fabry-Perot imaging of the [OIII]$\lambda5007$ line shows clearly defined 
ionization cones from the center of NGC~1068, inclined such that the disk 
of the galaxy is illuminated out to $r\sim11$~kpc 
\citep[Fig.~\ref{cartoon}, adapted from \citet{bruhweiler01} and 
\citet{veilleux03};][]{unger92, veilleux03}.
The [NII]$\lambda6548,6583$ line profile data of \citet{cecil90} 
showed that the kinematics of the ionized gas are well fit by a model 
in which the ionization cones intersect the plane of the galaxy, rather than 
illuminating halo gas seen in projection onto the disk.
The brightest region of [OIII] emission corresponds with the region of
peak radio continuum flux, extending $\sim10^{\prime\prime}$ NE of 
the nucleus \citep{condon90, morganti93}.  A knot of [OIII] emission is 
seen at the distance of the inner spiral, with further [OIII] emission 
extending several kpc further out into the disk of the galaxy.
Kinematics show that the gas in the ionization cones is ionized in situ 
by the central engine, 
and is not ejecta from the nuclear region.  Spectral line ratios also 
confirm that the AGN is the ionizing source even at $r\sim11$~kpc
\citep{veilleux03}, though the inner spiral itself shows line ratios 
typical of star forming regions.  The opening angle of the 
ionization cones is $40^{\circ}$, centered at position angle $31^{\circ}$
\citep{unger92}.
\citet{veilleux03} showed that the [OIII] emission is biconical, though 
much fainter to the SW than the NE.  This is consistent with the 
conclusion of \citet{cecil90} that the ionization cones are inclined with 
respect to the plane of the disk, illuminating the near side of the 
disk to the NE and the far side of the disk to the SW.
\citet{bruhweiler01} showed that the [OIII]/H$\beta$ ratio both interior
and exterior to the inner spiral along the cones is consistent with 
photoionization from 
the AGN without significant contribution from shocks.
\citet{usero04} detected the inner spiral in SiO, which is typically 
interpreted as a signature of shocks.  Infrared emission peaks have 
been detected along the ionization cones at the distance of the inner
spiral \citep{td88, lefloch01, marco03}.  \citet{bv93} and \citet{bv97} argued 
that the IR emission results from dust grains near the surface of molecular 
clouds heated by the X-ray and UV flux from the AGN.  Using ISO, 
\citet{lefloch01} detected the peaks in the $7.7~\mu$m PAH band, indicating 
that at least some of the emission is from PAHs rather than large dust grains.

Is it possible that the inner spiral is not illuminated by the AGN, but 
simply {\it appears} to be illuminated due to projection effects?  If the 
inner spiral lies in the plane of the disk, the AGN ionization cones
would have to be inclined well out of the plane of the disk and have a
small opening angle so as not to intersect it.  As noted 
previously, the [NII] and [OIII] data sets provide strong
evidence that the ionization cones intersect the disk.  
\citet{packham97} used near-IR polarimetry to produce a three dimensional 
model of the electron scattering cone, concluding that it is 
inclined by $43^{\circ}$ to the plane of the sky, corresponding to 
an inclination of $\leq10^{\circ}$ to the plane of the galaxy.  With 
an opening half-angle of $40^{\circ}$, it clearly intersects the disk.  The 
remaining possibility is that the inner spiral is inclined relative 
to the disk such that the inner spiral is not illuminated by the AGN.
The \citet{packham97} model requires an inclination of at least 
$30^{\circ}$ for the inner spiral to lie outside of the ionization cone.  Such 
an inclination is both physically implausible and would have been easily 
detected in kinematic data (e.~g. CO or H$\alpha$).  We therefore 
conclude that the AGN illuminates the inner spiral in the region of 
the PAH knots.  The fact that optical line ratios characteristic of AGN 
ionization are seen both immediately interior and exterior to the inner spiral
indicates that material in the inner spiral cannot be completely 
shielded from the AGN by intervening material.

In this paper we present the results from analysis of new infrared 
observations of NGC~1068 from the {\it Spitzer} Space Telescope.  
Observations with the IRAC \citep{iracref}, IRS \citep{irsref}, and MIPS 
\citep{mipsref} instruments are discussed in \S~2.  Analysis of the data 
is presented in \S~3, and conclusions are given in \S~4.  

\section{Observations and Data Analysis}

\subsection{Imaging}

NGC~1068 was observed with IRAC by the GTO program of Fazio et al.  
(PID \#32).  
MIPS observations were carried out as part of the
Great Observatories All-sky LIRG Survey
(GOALS, GO-1 RBGS LIRG/ULIRG program of Mazzarella et al.; PID \#3672).
The general data reduction procedures used on the IRAC and MIPS data were 
described in detail in \citet{mazz1}.  Here we discuss the special 
processing required for NGC~1068.  For reference, the bright star 
forming regions are at a radius of $15^{\prime\prime}$ (1~kpc).

Muxbleed saturation artifacts (see the IRAC Data Handbook\footnote{http://ssc.spitzer.caltech.edu/irac/dh/} for details)
in the $5.8~\mu$m and $8~\mu$m images presented challenges 
in the nuclear region of NGC~1068 within a radius of $15^{\prime\prime}$
(1 kpc).  A rough correction for this problem consisted
of measuring each artifact in a suitable aperture and
subtracting off the flux density of an identical aperture measured at
the same galaxy surface brightness level in a region outside
the artifact. This procedure resulted in a correction (reduction) of the 
flux density inside a $15^{\prime\prime}$ aperture by 
2.74 Jy (16\%) at 8.0~\micron\ and 0.41 Jy (5\%) at 5.8~\micron.

Since muxbleed is a result of saturation, an additional correction was
needed to account for the lost flux density.  Such saturation corrections 
were estimated by comparison with available data in the literature.
The flux density measured within a $6^{\prime\prime}$ diameter aperture was
$3.2 \pm 0.2$ Jy at $5.0~\mu$m \citep{rieke1}
and $25 \pm 3$ Jy at $10.0~\mu$m \citep{rieke2}.
A simple linear interpolation from these values
to the IRAC wavelengths predicts $6.2 \pm 0.5$ Jy and $16 \pm 2$ Jy
at 5.8 and $8.0~\mu$m, respectively.  The blue end of the transmission
window of the filter used by \citet{rieke2} extends to $7.6~\mu$m
\citep{low71}, so the $10~\mu$m measurement includes the majority of the 
flux density from the $7.7~\mu$m PAH feature.  
The {\it Spitzer} flux densities 
measured in a $6^{\prime\prime}$ diameter aperture centered on the nucleus 
of NGC 1068 are $7.9$ Jy at $5.8~\mu$m and $7.0$ Jy at $8.0~\mu$m.
(We note that the first muxbleed artifact in the $8~\mu$m PSF
is located at a radius of about $6^{\prime\prime}$, which is well outside 
this comparison region with a $6^{\prime\prime}$ {\it diameter}.)
The $5.8~\mu$m value measured from our corrected IRAC data
agrees with the estimate based on the Rieke \& Low measurements
within the uncertainties of our correction method.  However, since our 
$8~\mu$m IRAC estimate is 9 Jy below the estimate based on interpolation 
of the Rieke \& Low data, we have added 9.0 Jy to the $8~\mu$m
IRAC aperture measurements for NGC~1068 to account for the flux density 
lost to saturation. 

Photometry of the inner spiral was measured by taking the difference 
between measurements of $12^{\prime\prime}$ and $18^{\prime\prime}$ 
radius (800--1200~pc).  Extended source corrections were applied to 
each aperture measurement according to the equation derived by 
T.~Jarrett\footnote{http://spider.ipac.caltech.edu/staff/jarrett/irac/calibration/ext\_apercorr.html}.
The emission peaks on the inner spiral were measured using small 
elliptical apertures.  The wings of the PSF of the bright AGN would 
bias the observed flux densities if the emission peaks were measured 
relative to the sky background, so instead the peaks were measured 
relative to a local background elsewhere on the inner spiral.  In 
the case of the NE knot, the local background measurements were chosen 
to also lie along diffraction spikes at $3.6$ and $4.5\mu$m.

The MIPS $24~\mu$m image of NGC~1068 is dominated by a
bright unresolved central point source (the AGN) which saturated within
radius $\sim13^{\prime\prime}$.  To measure the spatially 
resolved features we subtracted a PSF model of the 
central point source.  The IDL package IDP3\footnote{http://mips.as.arizona.edu/mipspage/IDP3/idp3example.html}
was used to centroid the galaxy and PSF images; then the scale factor 
factor of the PSF image was interactively adjusted to best subtract off the 
point source component of the galaxy image.  As recommended by the 
MIPS Data Handbook\footnote{http://ssc.spitzer.caltech.edu/mips/dh/},
an object of similar color was chosen to create the PSF image.  The 
ULIRG IRAS~F05189-2524 served as a satisfactory PSF for this purpose, 
as it is both very bright and unresolved by {\it Spitzer} at $24\mu$m.  
IRAS~F05189-2524 is also a Seyfert 2 galaxy, and is the best match to 
the IRAS $60~\mu$m/$25~\mu$m 
color of NGC~1068, with a color of 3.82 compared to 2.24 for the latter.
The only galaxies in the GOALS sample with colors more closely matching 
NGC~1068 were spatially resolved in the MIPS $24~\mu$m images.
Contours of the PSF-subtracted $24~\mu$m image of NGC~1068 are overlaid 
on the $8~\mu$m 
image in Fig.~\ref{n1068i4m1}.  The elongation of the masked region on 
the $8~\mu$m image is due to the muxbleed residuals discussed above.

The best fit PSF for NGC~1068 was also used to estimate the total 24 \micron\ 
flux density, which could not be measured directly due to saturation.
The point-source component of the total flux density was estimated by 
taking the flux density of IRAS~F05189-2524 (3.04~Jy) and multiplying by 
the best fitting PSF scale factor (25), resulting in an estimate of 75.90~Jy.
In addition, the extended 24 \micron\ emission from NGC~1068 was measured 
directly from the PSF-subtracted image.  Combining the 0.5~Jy extended 
component with the point-source component yields an estimated total 
24 \micron\ flux density of NGC~1068 (before aperture and color corrections) 
of 76.4~Jy.  The uncertainty in the empirically determined PSF scale factor 
(25.0) is about 20\%, and this dominates the uncertainty in this
24 \micron\ total flux density measurement.

Table~1 presents the {\it Spitzer} flux density measurements and estimated 
uncertainties.  As described in detail in 
\citet{mazz1}, MIPS color corrections have been applied based on the 
IRAS and MIPS SEDs.  With aperture and color corrections, the
total estimated 24~\micron\ flux density using the {\it {\it Spitzer}}
data is 79.6 Jy. Comparison with the IRAS $25~\mu$m  measurements of
$87.6 \pm 0.2$ Jy \citep{rbgs} indicates that
this $24~\mu$m estimate from {\it {\it Spitzer}} is low by
about 9\%. This is consistent with the systematic mean
offset of 20\% (with a $1\sigma$ scatter of 15\%)
observed between the $24~\mu$m {\it {\it Spitzer}} measurements and
$25~\mu$m IRAS measurements across our entire sample of 203 systems 
\citep{mazz1}.  We are therefore confident in the PSF fitting procedure, 
from which it can be stated that more than 99\% (75.9/76.4) of the $24~\mu$m
emission from NGC~1068 originates within a radius of
$\approx30^{\prime\prime}$ (the first Airy ring).  Approximately 80\% of
the $24~\mu$m emission originates within the saturated region 
$r<13^{\prime\prime}$.
NGC~1068 also saturated the $160~\mu$m array and a flux density estimate could
not be recovered. 

\subsection{Spectroscopy}

The nucleus of the galaxy was observed at high spectral 
resolution by the GTO program of Houck et al. (PID \#14), and mapped at 
low resolution by the GO-1 program of Gallimore et al. (PID \#3269).
The Short-Low (SL) IRS maps consisted of 13 half slit-width  
(1.8 arcsec) steps for the first and second order slits.  The maps 
were centered on the nucleus.  There were no dedicated off (sky) 
observations, but a local background was subtracted by using
the non-primary slit, which is centered approximately 80 arcsec  
from the nucleus, for both the SL1 and SL2 maps.  The IRAC $8~\mu$m surface 
brightness at the locations selected for the background are typically 
$10^{-4}$ -- $10^{-3}$ of the level at the nucleus.
CUBISM\footnote{http://turtle.as.arizona.edu/jdsmith/cubism.php} was 
used to perform the local background subtraction and form the SL spectral 
cubes.  From these cubes we were able to identify regions
over which to extract 1D spectra, and we used CUBISM to perform  
this extraction in matched apertures in SL1 and SL2.  In addition, we 
used CUBISM to generate a pure PAH map over the area covered by the IRS slits.
Aperture locations and the PAH map are shown in Fig.~\ref{regions}.  
Fluxes and equivalent widths of the low resolution data were measured 
using PAHFIT \citep{pahfit}.
The regions closest to the nucleus suffered from saturation and did 
not yield usable spectra; see Fig.~\ref{regions}, lower panel.  
Table~2 presents the equivalent widths and line 
fluxes of the emission features measured in each aperture.

The Short-High and Long-High IRS observations do not suffer from saturation, 
but only cover a single position at the center of the galaxy.  No background 
subtraction was performed due to the lack of a background slit observation.
Spectra were extracted using 
SPICE\footnote{http://ssc.spitzer.caltech.edu/postbcd/spice.html}, and 
lines were measured on the extracted spectra using SMART \citep{smartref}.  
Our high resolution spectral line measurements are presented in Table~3.

\subsection{Separating Non-Stellar from Stellar Emission in Seyfert Galaxies}

Although IRS spectra provide the ideal measure of the presence and strength 
of PAH emission, the available data cover only a portion of the inner spiral.
IRAC and MIPS imaging is required to investigate the region outside 
of the spectral map.  There are several possible sources for the IR emission 
from Seyfert galaxies.  The power-law 
continuum from an AGN can contribute to all IRAC wavelengths.  The 
$3.6~\mu$m and $4.5~\mu$m data are dominated by the Rayleigh-Jeans 
tail of stellar photospheric emission.  The $5.8~\mu$m channel covers the
wavelength range which includes the $6.2~\mu$m PAH line, while the 
$8~\mu$m channel includes the $7.7~\mu$m 
and $8.6~\mu$m PAH emission features.  Hot dust can also be a significant 
contribution at all IRAC wavelengths.  Stellar continuum emission is 
insignificant compared to PAH emission in the IRAC $8~\mu$m band
in nearly all nearby LIRGs \citep{mazz1}.  The stellar continuum flux density 
at $8~\mu$m is estimated to be 19--23\% of the $3.6~\mu$m flux density
\citep{pahre04, dalesings}; based on this we estimate that stars contribute 
only a few percent
(for example, NGC~1068 1~kpc aperture: $2.8\times0.23/17 = 0.038$) of the 
$8~\mu$m flux densities of this galaxy.  To test for spatial variations 
between the starlight and PAH emission in NGC~1068, a scaled $3.6~\mu$m 
image was subtracted from the $8~\mu$m image.  The ratio of this nonstellar 
emission image to 
the $8~\mu$m image is constant with little variation across the galaxy 
(excluding the central region which saturated at $8~\mu$m, as noted previously),
showing that starlight accounts for $\sim4~\%$ of the $8~\mu$m emission 
throughout the disk of NGC~1068.

\begin{figure}
\vbox{
\begin{center}
\includegraphics[width=\textwidth]{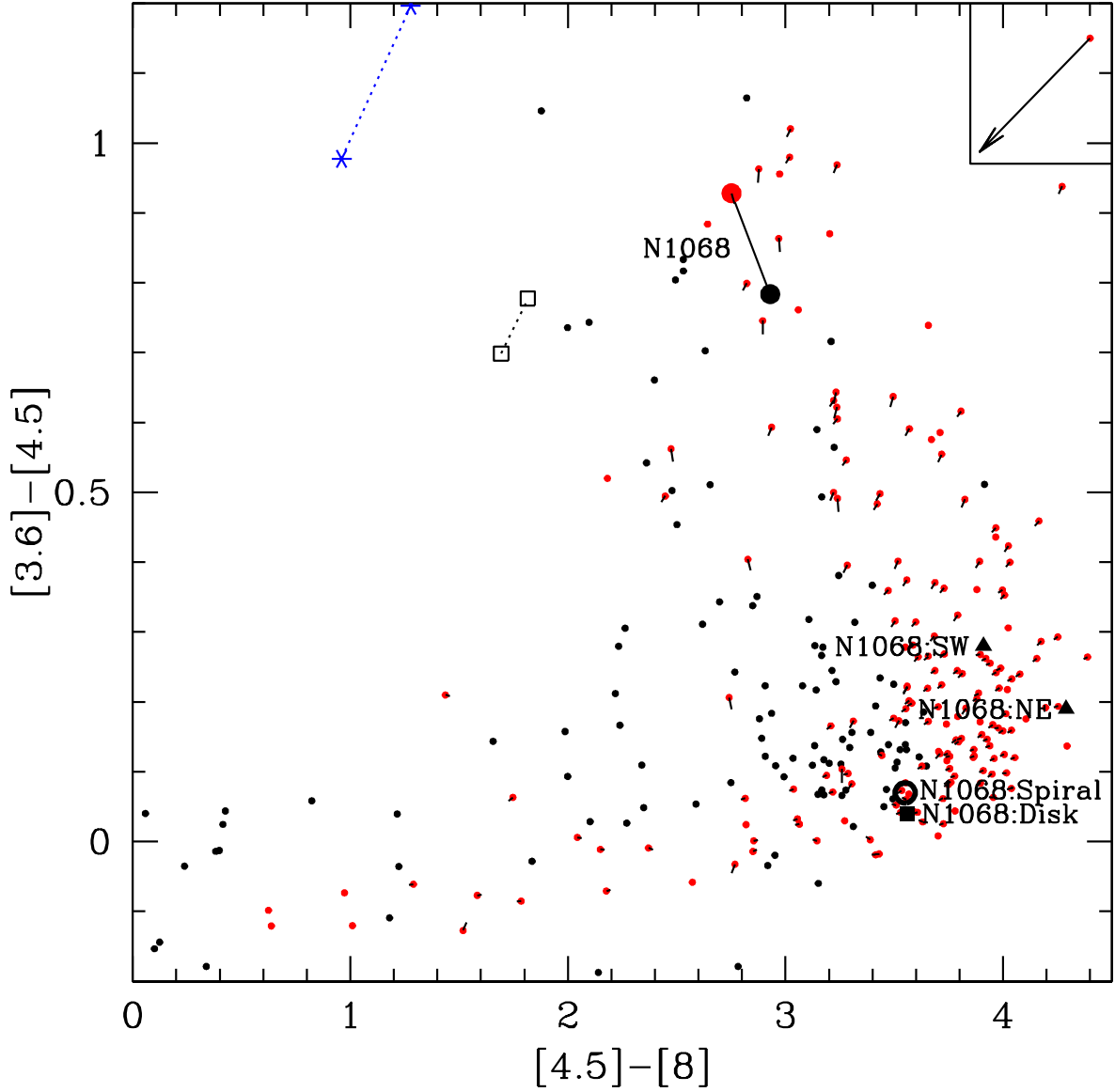}
\figcaption{\small
IRAC colors of the RBGS LIRGs and ULIRGs from \citet{mazz1}.  NGC~1068 
is marked (1~kpc radius and total apertures; red and black points 
respectively), as are the colors of the two bright knots on the inner spiral
and the colors of the galaxy disk ($r>1$~kpc) and the inner spiral.
Small red points indicate the central 1~kpc radius aperture measurements 
of the rest of the GOALS sample, 
while black points are total magnitudes.  Unit vectors indicating the 
change in color from the central aperture to the total aperture are shown; 
note that there is not a one to one correspondence between central and 
total aperture measurements due to a combination of resolution and 
multiple interacting galaxies, respectively.  The median change in 
color from the central 1~kpc aperture to the total aperture is shown 
by the arrow in the upper right.  Colors 
are expressed in the Vega magnitude system, so by definition a pure 
starlight SED would appear near zero in both color axes.  
Also shown are the colors of a pure AGN power-law spectrum (open squares, 
from lower left to upper right: $\alpha = 0.84$, \citet{clavel}; 
$\alpha = 1.4$, \citet{neugebauer}), and hot dust (blue points for 
1000~K and 900~K).  
\label{cc}
}
\end{center}}
\end{figure}

Determining the presence of PAHs using IRAC and MIPS photometry 
requires the measurement and modeling of the contribution of AGN continuum 
and hot dust at $8~\mu$m.  Several 
color-color diagnostics have been used to distinguish AGN-dominated 
systems and starburst-dominated systems in the literature 
\citep[e.~g.][]{stern05}.  The GOALS sample \citep{mazz1} is shown in 
the [3.6]-[4.5] vs. [4.5]-[8] color-color space in Fig.~\ref{cc}.  Also 
shown are the colors of hot dust and 
AGN power laws.  Since the vertical axis is the ratio of 
two channels normally dominated by starlight, galaxies without significant 
AGN or hot dust lie near zero.  The horizontal axis separates galaxies with 
little or no PAH emission (left) from those dominated by PAH emission 
(right).  The [4.5]-[8] color measures the jump from the Rayleigh-Jeans 
tail of starlight at $4.5~\mu$m (plus AGN continuum, if present) to the 
bright PAH emission at $8~\mu$m.  The better-known Stern~et~al. color-color 
space uses [5.8]-[8] on the horizontal axis instead of [4.5]-[8]. 
Since both the 5.8 and $8~\mu$m channels include PAH bands, [4.5]-[8]
is preferred due to the clearer physical interpretation.

Combinations of IRAC images can be used to create spatial maps of different 
components of IR emission.  \citet{pahre04} used 3.6 and $4.5~\mu$m data to 
estimate the contribution of starlight, which was then scaled down in 
proportion to a Rayleigh-Jeans tail and subtracted from the $8~\mu$m image
to create a map of nonstellar (typically PAH) emission.  As noted above, 
this technique has little effect on LIRG and ULIRG data such as those 
examined here since the $8~\mu$m images are dominated by nonstellar emission. 
A similar method can be used to identify nonstellar emission at $4.5~\mu$m.
No PAH emission lines are present at such short wavelengths, 
so the nonstellar emission sources are direct radiation from the AGN and 
hot dust only.  Making an 
assumption about the form of the nonstellar emission, this technique can 
be used to quantitatively estimate the emission from stars and the assumed 
nonstellar source at 3.6 and $4.5~\mu$m.  The ratio of stellar emission at 
$3.6~\mu$m to that at $4.5~\mu$m, $f_*$, is known; using \citet{pahre04} 
$f_*=1.7$.  A similar ratio, $f_{\rm pl}$ is easily calculated for any given 
power law slope $\alpha$.  For typical values 
$0.84<\alpha<1.4$, $0.83<f_{\rm pl}<0.73$.  Image subtraction is then 
performed, with the power law component of the $4.5~\mu$m image given by 
$I_{\rm pl}(4.5) = \frac{{f_*}I(4.5)-I(3.6)}{{f_*}-f_{\rm pl}}$, where 
$I(\lambda)$ is the IRAC image at a given wavelength.  Similar equations 
are easily derived for the stellar components of each image, or the power 
law component of the $3.6~\mu$m image.  Note that if the nonstellar emission 
arises from hot dust this method will only reveal the locations of such 
emission and one can simply use $I(4.5)-I(3.6)/f_*$ without attempting to 
account for the nonstellar contribution to the $3.6~\mu$m flux 
\citep{engel05}.  A quantitative flux density estimate would require an 
assumed dust temperature from which $f_{\rm dust}$ could be calculated, 
replacing $f_{\rm pl}$ in the above prescription.  In either case, the assumed 
nonstellar emission can also be extended to $8~\mu$m to better constrain 
the PAH emission.

\section{Discussion}

\subsection{Spitzer Data}

The IRAC colors of the region within 1~kpc of the nucleus and the total 
emission from the galaxy place NGC~1068 in the AGN region of the IRAC 
color-color diagram defined by \citet{stern05}.
Emission at IRAC wavelengths is highly concentrated, with 70--80\% of 
the flux density coming from the central kpc (Table~1).

High resolution spectra of the nucleus show strong AGN signatures in a 
number of diagnostic line ratios.  Line fluxes and equivalent widths are 
presented in Table~3. \\ 
The [O~IV]$25.9~\mu$m/[Ne~II]$12.8\mu$m line ratio is used as a 
diagnostic for the fractional contribution of the AGN to the luminosity of 
a galaxy \citep{genzel98,armus}.  The NGC~1068 nucleus has a ratio of 4.2 as 
measured from the IRS spectra, compared to 2.7 as measured from the ISO 
data of \citet{sturm02}.  
[Ne~V]$14.3~\mu$m/[Ne~II]$12.8\mu$m is a similar diagnostic ratio; 
the nucleus of NGC~1068 has a ratio of 2.1, compared to the ISO ratio of 1.4.
Although these AGN diagnostic ratios are higher than the ISO measurements 
of \citet{sturm02}, this is not surprising since the IRS slits are 
smaller than the corresponding ISO aperture by roughly a factor of two.
With less light from the disk of the galaxy, the line ratios might be expected
to show higher ionization.  Using the [Ne~V]$14.3~\mu$m/[Ne~III]$15.6~\mu$m 
and [Ne~II]$12.8\mu$m/[Ne~III]$15.6~\mu$m line ratios measured using the high 
resolution spectrum, the diagnostic diagram of \citet{voit92a} indicates an 
ionization parameter ${\rm log}~U\sim-2$ and a power law input spectrum of 
index $\alpha\sim-1.3$.  Compared to nearby 
AGNs, starbursts, and ULIRGs \citep{armus}, the line ratios in the 
nucleus of NGC~1068 are in excellent agreement with the AGN sample,
consistent with a 100\% AGN contribution to the observed mid-IR nuclear 
spectrum.  
 
\begin{figure}
\vbox{
\begin{center}
\includegraphics[angle=-90,width=\textwidth]{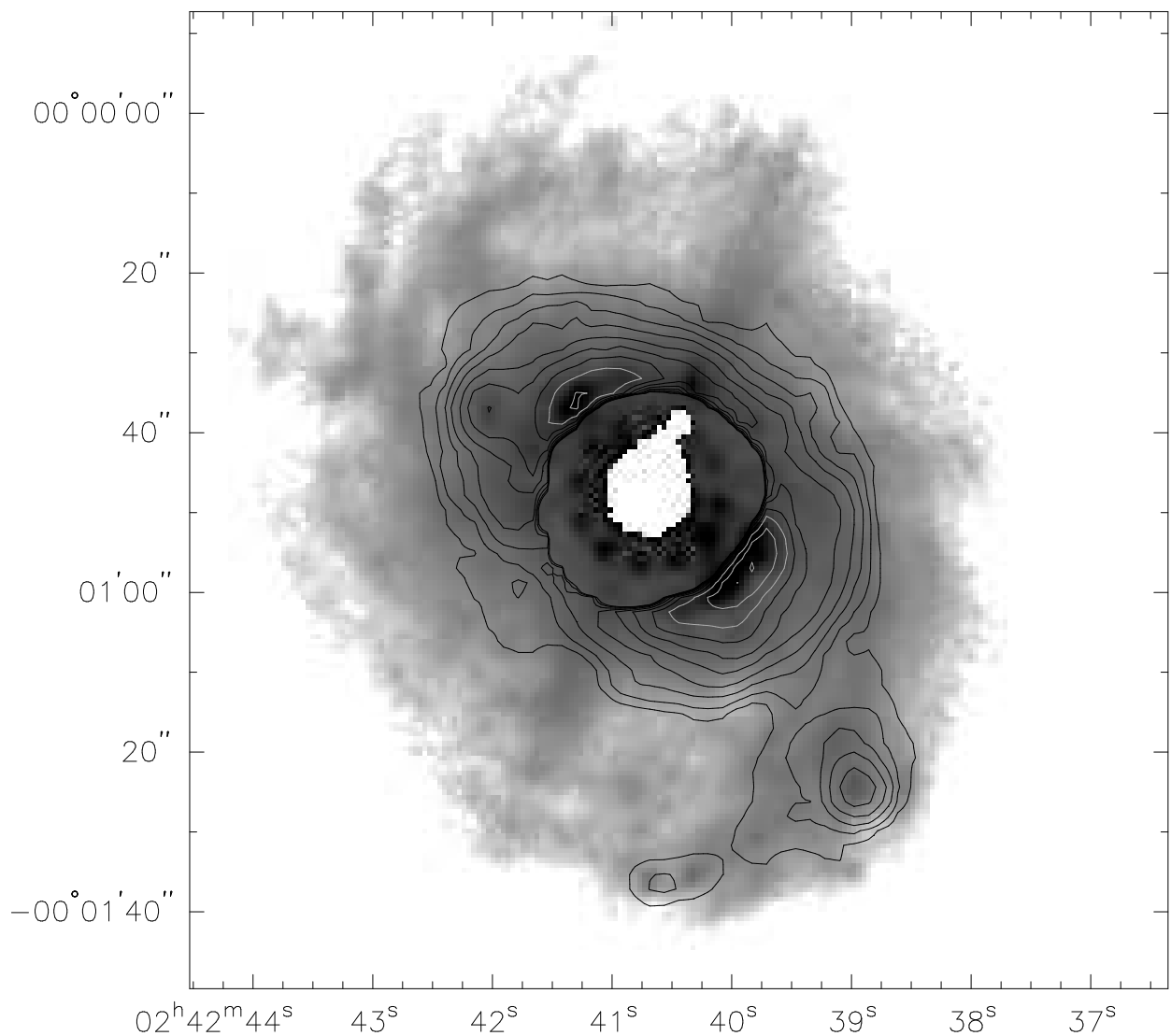}
\figcaption{\small
The $8~\mu$m image of NGC~1068 (grayscale), with contours of the PSF-subtracted
$24~\mu$m image.  The brightest contour levels are shown in light gray for 
clarity.  The saturated data in the nucleus have been excluded from 
both images.  Residual banding is visible on the $8~\mu$m image as the streak 
running SE-NW.  The ring of spots is also an artifact of the $8~\mu$m PSF.
The $24~\mu$m data show many of the features seen at $8~\mu$m, including the 
two knots on the inner spiral, the NE tip of the southern arm of the 
inner spiral 1.5~kpc ($23^{\prime\prime}$) ENE of the nucleus, and the 
peak 3~kpc ($45^{\prime\prime}$) SW of the nucleus.
\label{n1068i4m1}
}
\end{center}}
\end{figure}

\begin{figure}
\vbox{
\begin{center}
\includegraphics[width=2.5in]{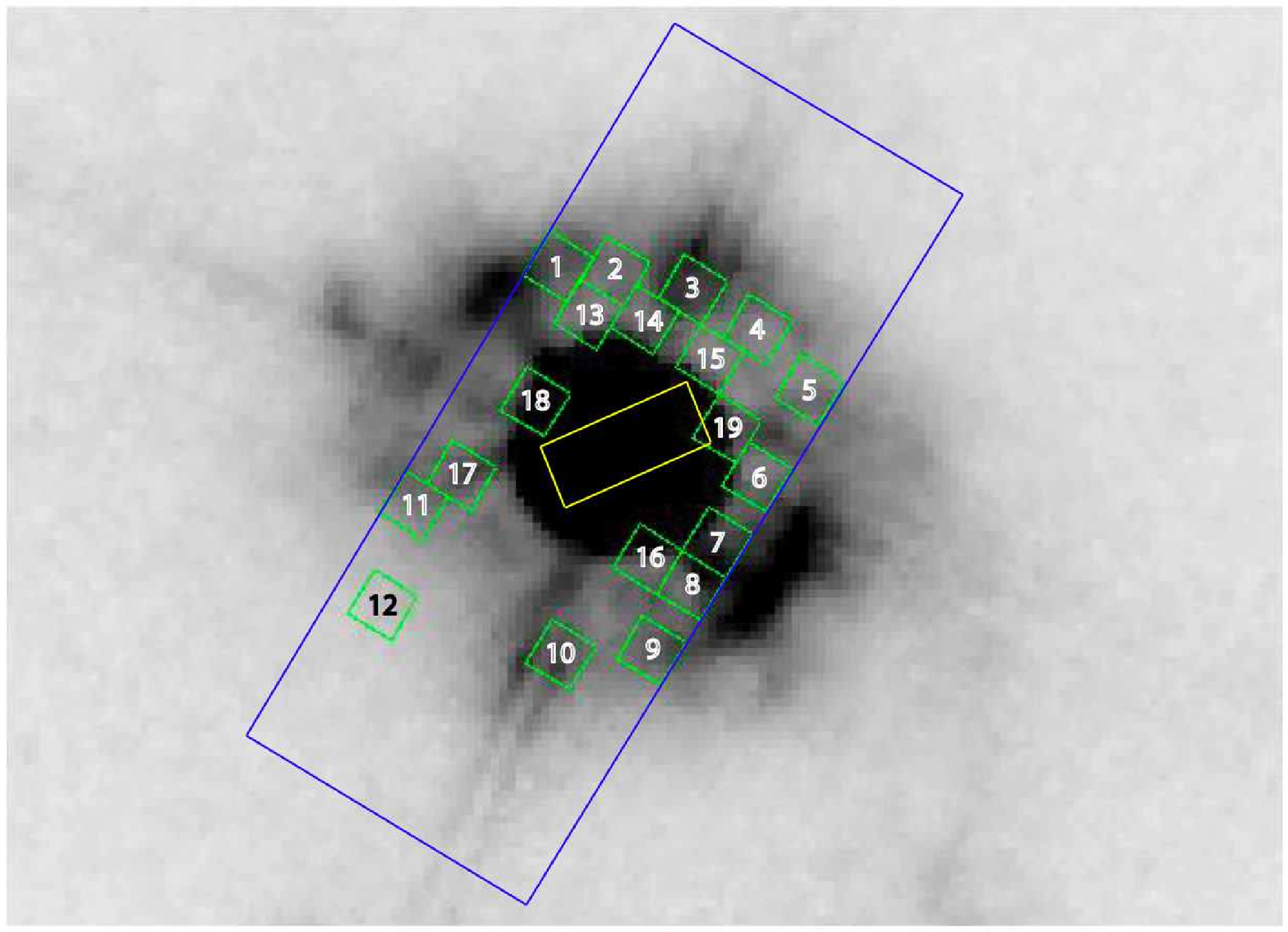}\\
\includegraphics[width=4in]{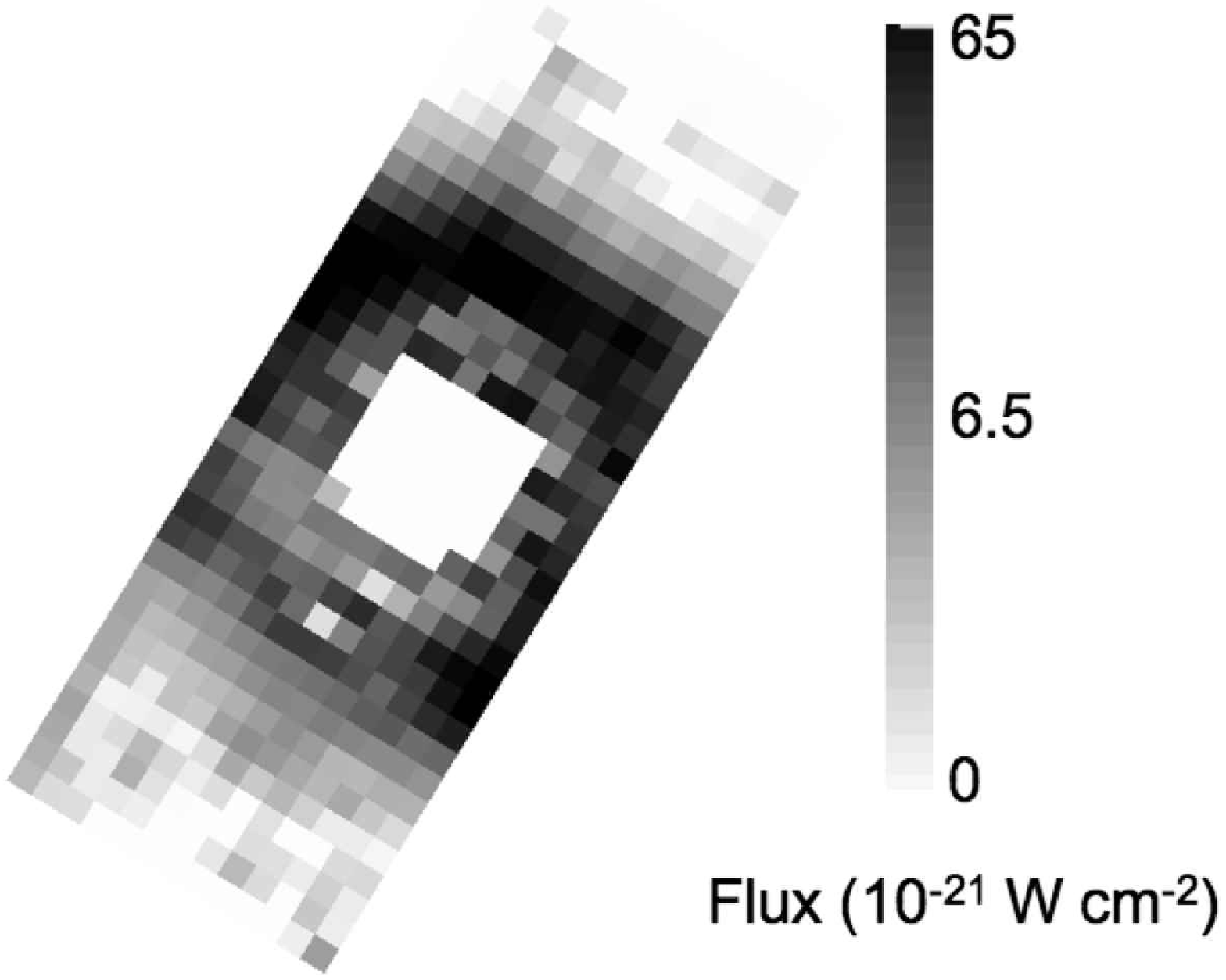}
\figcaption{\small
Extraction apertures for the Short-Low IRS spectra are overlaid on the 
$5.8~\mu$m IRAC mosaic (top, green numbered boxes).  Also shown are the 
area covered by the Short-Low map (blue box) and the position of the 
Short-High slit (yellow box).  Banding and PSF artifacts are as described 
in Fig.~\ref{n1068i4m1}.  North is up and east is to the left.
The map of the $6.2~\mu$m PAH feature is shown in the bottom panel.
The center of the galaxy has been masked out.
\label{regions}
}
\end{center}}
\end{figure}

The IRAC images show bright $8~\mu$m peaks (also detected in the other 
IRAC channels) along the NE and SW arcs at ${\rm PA}\sim45^{\circ}$ in 
the inner spiral (Fig.~\ref{n1068i4m1}) at the 
locations first detected by \citet{td88}.  Of the total $8~\mu$m
flux density within the annulus $800 < R < 1200$~pc from the nucleus, 
52\% is emitted by these two knots.  The surface brightness of the knots 
(400--500~MJy/sr) is more than twice that seen elsewhere in the inner spiral
($\sim200$~MJy/sr).  Compared to the central kpc of the galaxy,
these knots are bluer by more than half a magnitude in [3.6]-[4.5] and redder 
by 0.55-0.85~mag in [4.5]-[8].  Since the wavelength range of the $8~\mu$m 
channel includes both the $7.7~\mu$m and $8.6~\mu$m PAH features,
the latter color strongly suggests that the 
knots are dominated by PAH emission.  

\begin{figure}
\vbox{
\begin{center}
\includegraphics[width=\textwidth]{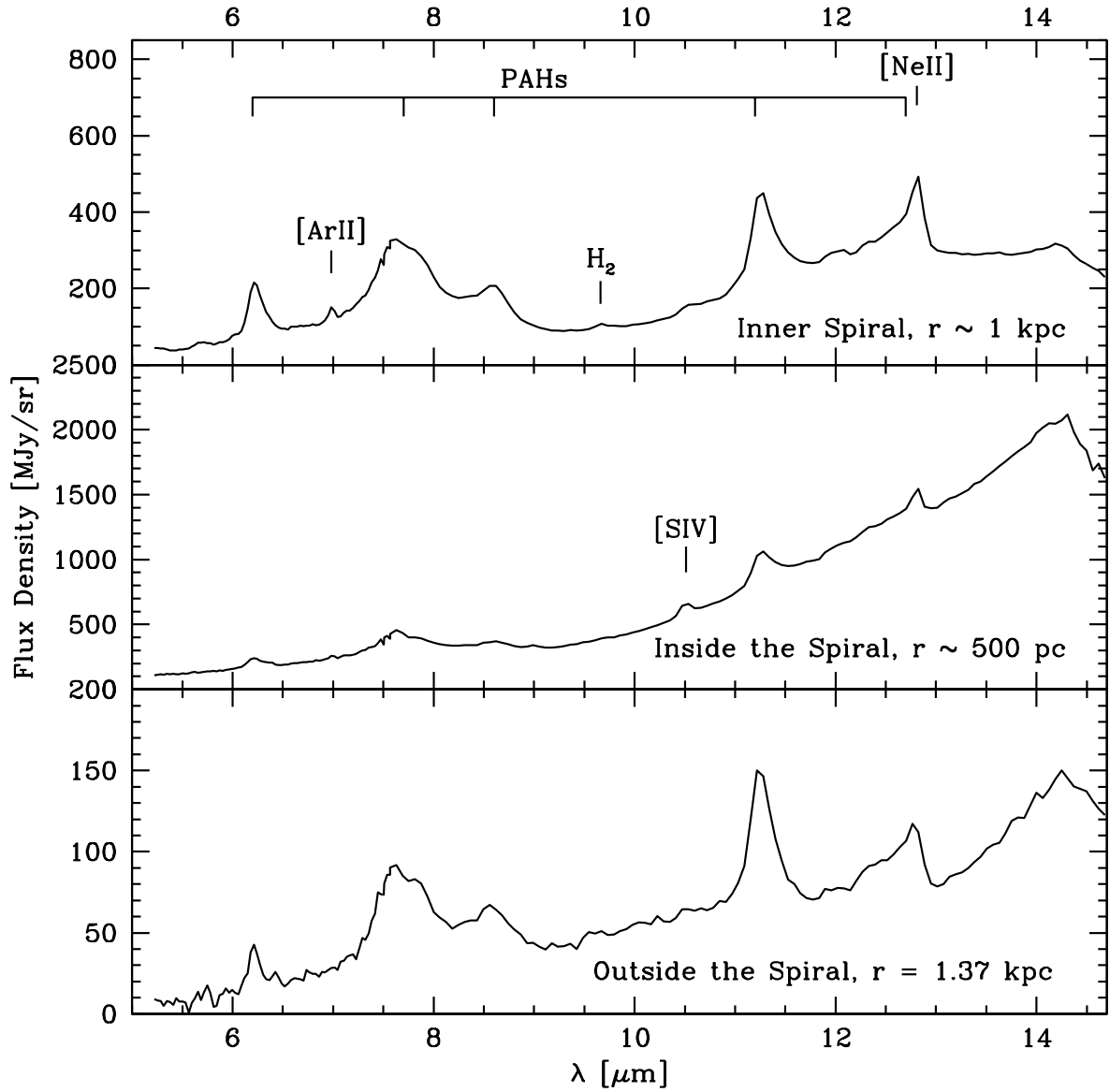}
\figcaption{\small
IRS Short-Low spectra of NGC~1068.  The top panel shows the average of 
eight apertures on the inner spiral, the middle panel shows the average 
of the three apertures nearest the nucleus, and the bottom panel shows 
aperture~12, outside the inner spiral.
\label{pahspec}
}
\end{center}}
\end{figure}

The low resolution IRS data confirm the predominance of PAH emission in 
the knots observed in the IRAC $8\mu$m  image.  The $6.2~\mu$m, $7.7~\mu$m, 
and $11.2~\mu$m PAH features, 
plus the PAH$12.7\mu$m$+[{\rm Ne~II}]12.8\mu$m blend, are clearly visible 
throughout the inner spiral.  Figure~\ref{pahspec} shows representative 
spectra, including the inner spiral, the region interior to the inner spiral, 
and the region outside of the inner spiral.  Fluxes and equivalent widths 
of each measured emission feature are presented in Table~2.  The PAH 
features are strongest in the inner spiral (apertures 1--5, 9--11; top 
panel of Fig.~\ref{pahspec}).  However, such features are also 
present in the innermost usable apertures (apertures 16, 18, \& 19; 
middle panel of Fig.~\ref{pahspec}).  Although the PAH equivalent widths 
in these innermost apertures are small due to the bright hot dust 
continuum near the AGN, the fluxes of the PAH lines are several times as 
bright as in aperture~12 (outside the inner spiral) and typically half 
as bright as in the inner spiral.  These innermost apertures also show the 
strongest [S~IV]~$10.51~\mu$m emission, while [Ar~II]~$6.98~\mu$m and 
${\rm H}_2~9.66~\mu$m emission are strongest on the inner spiral.

For comparison, the average starburst galaxy spectrum from \citet{brandl06}
was also analyzed using PAHFIT.  Line fluxes and equivalent widths 
measured using PAHFIT cannot be directly compared with measurements made 
using other techniques because PAHFIT takes into account the flux in the 
extended wings of the spectral lines; see \citet{pahfit} for details.
PAH equivalent widths are 
$2.36~\mu$m, $7.15~\mu$m, $1.09~\mu$m, $1.36~\mu$m, and $0.97~\mu$m
(6.2, 7.7, 8.6, 11.3, and $12.7~\mu$m features, respectively).  Though
the inner spiral is the source of the strongest PAH emission in NGC~1068,
this emission is significantly weaker than that seen in starburst galaxies.

\begin{figure}
\vbox{
\begin{center}
\includegraphics[width=\textwidth]{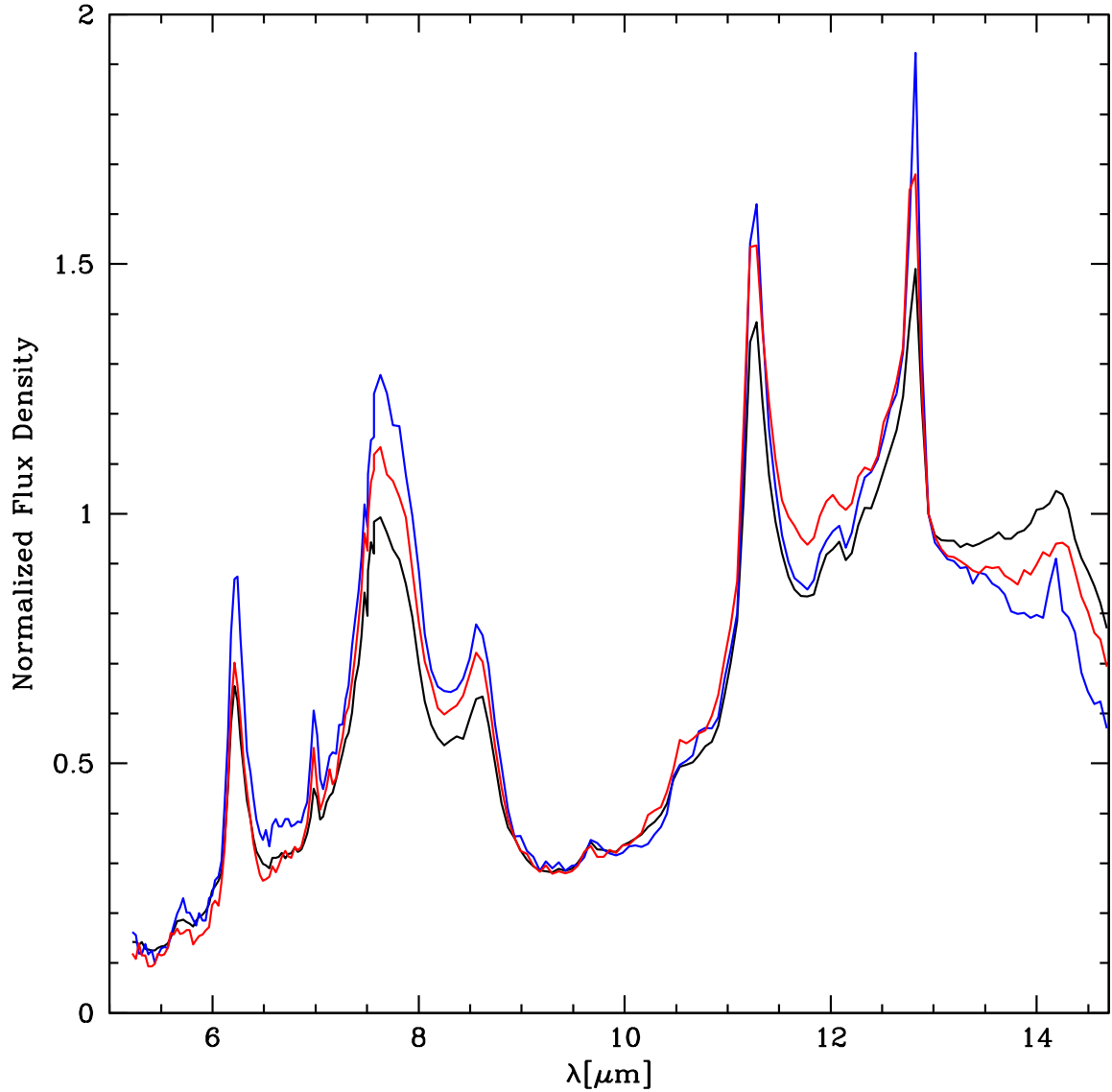}
\figcaption{\small
Low resolution spectra of the inner spiral within the ionization cones
(aperture~1, blue; aperture~9, red) are compared with the average spectrum 
of the inner spiral outside the ionization cones (black).  The three 
spectra have been normalized to a common flux density at $13~\mu$m.  No 
difference is found between the continuum flux density within the ionization 
cones compared to outside the ionization cones.
\label{ringspec}
}
\end{center}}
\end{figure}

With the detection of bright PAH emission lines throughout the inner spiral,
the model of \citet{bv93} in which the IR peaks on the inner spiral
originate from dust grains heated by the AGN can be ruled out.  A composite 
model in which the IR peaks are due to a combination of PAH emission plus 
continuum from hot dust heated by the AGN can also be strongly
constrained.  As shown in Fig.~\ref{regions}, apertures~1 and 9 both lie 
within the ionization cones and are the closest apertures to the PAH knots 
on the inner spiral.  If hot dust made a significant contribution to 
the flux in these apertures, the PAH equivalent widths would be lower 
than in apertures on regions of the inner spiral falling outside the 
ionization cones (apertures~3--5, 10, and 11).  Instead, the PAH equivalent 
widths measured in apertures~1 and 9 are as high or higher than those 
measured elsewhere on the inner spiral.  Any contribution of continuum 
flux from hot dust heated by the AGN must be too small to significantly 
affect the PAH equivalent widths.  We estimate the amount of additional 
continuum flux within the ionization cones compared to the rest of the 
inner spiral by comparing the flux ratios and equivalent width ratios 
of the PAH features.  PAH fluxes in apertures~1 and 9 are brighter than 
the average values in the rest of the inner spiral by factors of 20--80\%;
equivalent widths are brighter by factors of 10--100\%.  The signature of
a greater contribution from the continuum is a larger increase 
in flux than in equivalent width.  This is seen in aperture~9, with a 
flux increase of 31\% and equivalent width increase of $15\%$ in 
the $11.3\mu$m and $12.7\mu$m features that would be most affected by hot 
dust continuum.  These ratios indicate a continuum increase of $1.31/1.15=14\%$
compared to the average value in the rest of the inner spiral.  However, 
in aperture~1 the increase in PAH equivalent width is greater than or equal 
to the increase in PAH flux for every feature, indicating a continuum 
level equal to or less than that seen in the rest of the inner spiral.
Comparing the aperture~1 and 9 spectra to the average spectrum of the 
inner spiral apertures outside the ionization cones confirms the 
conclusion that little or no excess continuum is seen within the 
ionization cones.  Figure~\ref{ringspec} shows the three spectra normalized 
to a common flux density at $13~\mu$m.  The apertures within the ionization 
cones clearly show brighter PAH features than the average spectrum, 
while the continuum shows very little variation between the three spectra.
Warm dust visible at $24~\mu$m 
(Fig.~\ref{n1068i4m1}) is present at the locations of the peaks on the 
inner spiral, though it is unclear whether the AGN or local star formation 
activity is the primary heating source.

\begin{figure}
\vbox{
\begin{center}
\includegraphics[width=\textwidth]{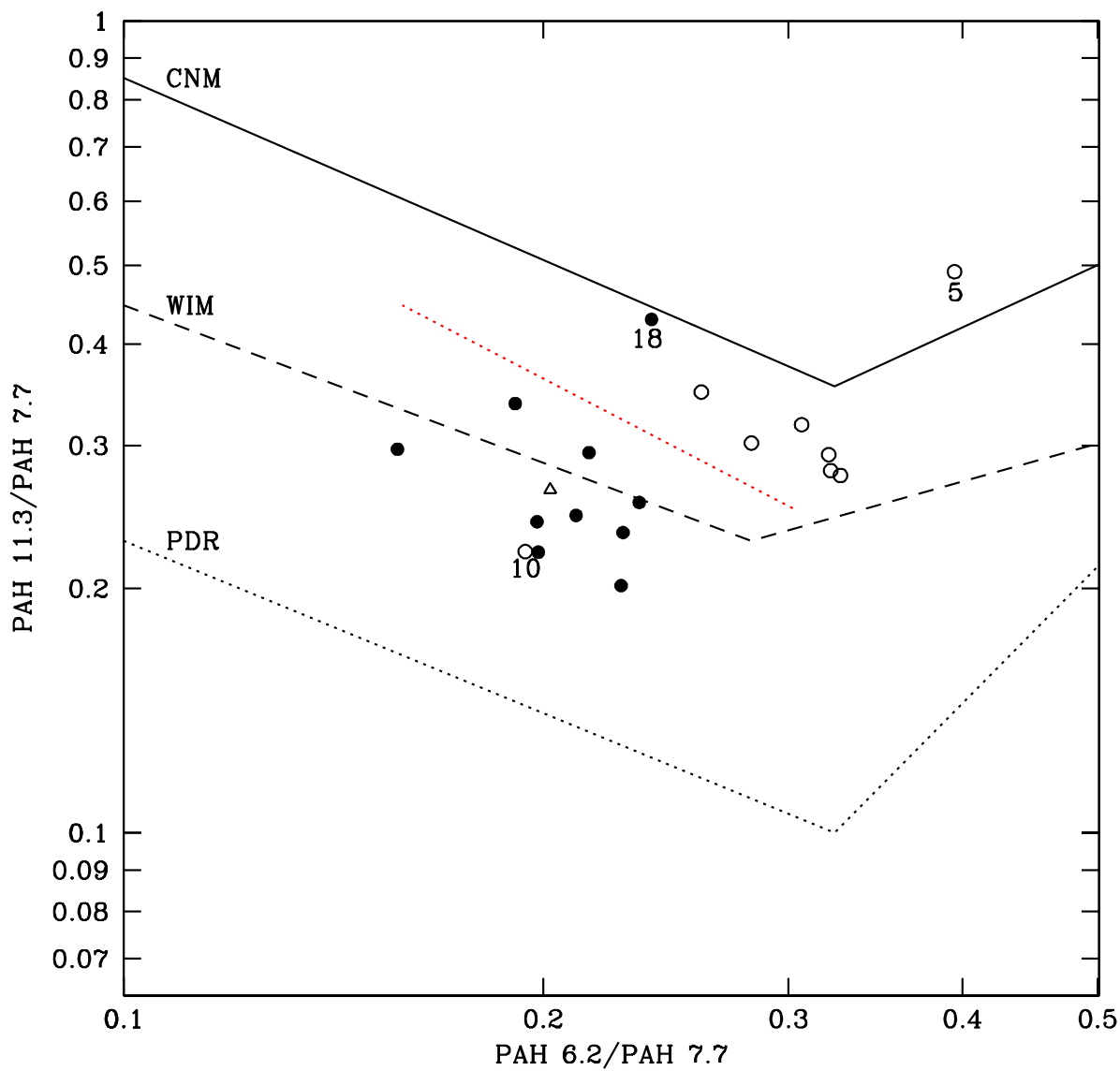}
\figcaption{\small
Relative strengths of three PAH emission features.  Apertures along the 
inner spiral are shown as open circles.  Apertures inside
the inner spiral are shown as solid circles 
The aperture exterior to the inner spiral is shown as an open triangle.
Outliers have been labeled by aperture number.
Uncertainties are $\sim0.01$ in each PAH flux ratio, and are not shown.
The lines are adapted from \citet{draineli01}.  The solid line 
corresponds to the CNM model, the dashed line corresponds to the WIM 
model, and the dotted line corresponds to the PDR model.  For all three 
models, PAH molecule size increases to the left.  The red dotted line 
shows part of the boundary of the region above which a simple ISM model 
can reproduce observed PAH features.  Data below this line require greater 
ionization.
\label{draine}
}
\end{center}}
\end{figure}

Ratios between the $6.2~\mu$m, $7.7~\mu$m, and $11.3~\mu$m PAH features
are shown in Fig.~\ref{draine}.  Lines adapted from \citet{draineli01}
show the locations of PAH emission from a cold neutral medium (CNM) model, 
a warm ionized medium (WIM) model, and a photodissociation region (PDR) 
model.  Also shown is a portion of the boundary of the region of PAH line 
ratios which can be reproduced by the composite ISM model of 
\citet{draineli01}.  With the exception of aperture~10, the PAH measurements 
of the inner spiral all lie above the boundary, within the region 
consistent with a composite model.
Aperture~5 is the outlier in the direction of a colder environment and 
smaller PAH molecules.  The ${\rm H}_2~9.66\mu$m line is nearly a factor of 
two brighter in this aperture than in the other apertures.
Interestingly, aperture~18 --- only 480~pc from the 
nucleus --- also lies in this cold ISM region of the diagram.  It is also 
interesting that aperture~10 on the inner spiral and aperture~12 outside 
the inner spiral both lie in the region requiring more ionizing radiation 
than the composite ISM model, with similar PAH ratios to the majority of the 
apertures interior to the inner spiral.  Since apertures 10 and 12 are 
both well 
outside the [OIII] ionization cone, ionization from the AGN is unlikely.  
Aperture~10 coincides with the location of the peak CO emission in the map 
of \citet{schinnerer}, suggesting the formation of massive stars as a likely 
ionization source.  The general picture presented by the ratios of PAH 
features is that PAH molecules in the inner spiral are smaller and lie 
in a cooler, less ionized environment than PAHs closer to the nucleus.

The CNM-like PAH ratios of aperture~18 suggest that radiation from the 
AGN may be absorbed within a few hundred parsecs along the NE ionization 
cone.  The radio jet of \citet{condon90} ends abruptly near the same 
location.  However, optical line ratios also show that radiation from the 
AGN is ionizing gas up to 11~kpc from the nucleus within the NE cone 
\citep[][\S~3.2]{veilleux03}.

\begin{figure}
\vbox{
\begin{center}
\mbox{\includegraphics[width=3.5in]{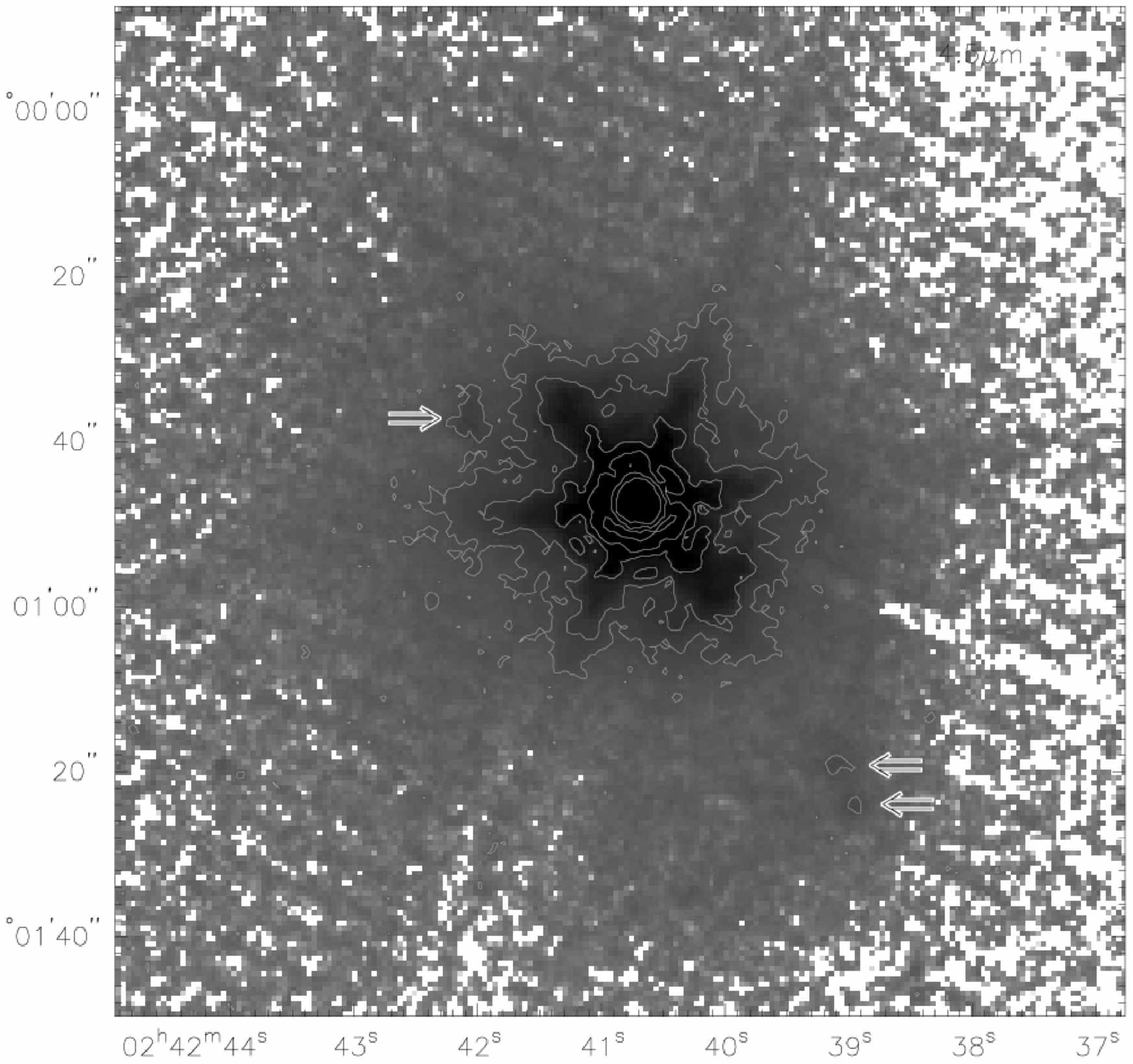}\quad
	\includegraphics[width=3.1in]{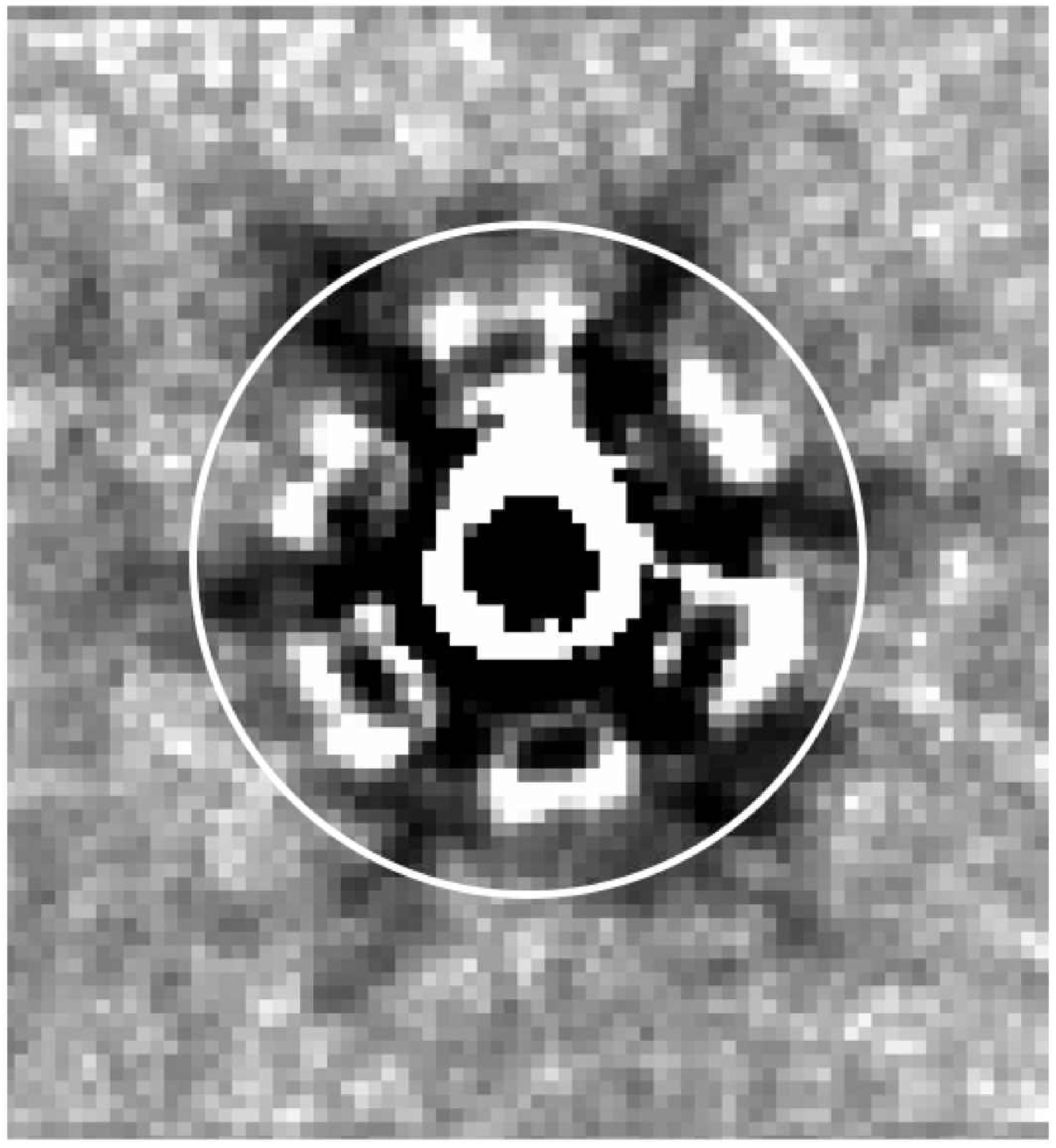}}
\figcaption{\small
Left: Contours of an estimate of the nonstellar emission overlaid on $4.5~\mu$m 
grayscale image of NGC~1068.  As expected, the AGN dominates the nonstellar 
emission, but the PAH emission regions outside the inner spiral (SW and ENE
of the nucleus; marked) are also visible.  The method used 
to estimate the nonstellar component of the $4.5~\mu$m flux density is 
described in \S~2; $\alpha=1$ was used as the power law slope.  
Right:  The $4.5\mu$m nonstellar emission from the point-like nucleus has 
been removed by PSF subtraction.  The circle marks the location 
of the inner spiral 1~kpc ($15^{\prime\prime}$) from the nucleus.  This 
image shows that there is 
very little detectable emission from hot dust outside of the nucleus.
\label{n1068ns}
}
\end{center}}
\end{figure}

Although the $24~\mu$m MIPS data are saturated ($r < 13^{\prime\prime}$, 
900~pc), the innermost available isophotes are elongated along the same NE-SW 
axis corresponding to the arcs seen at other wavelengths 
(Fig.~\ref{n1068i4m1}).  
In addition to the PAH knots, a weaker feature is visible in both the 
IRAC and MIPS~$24~\mu$m images, corresponding to the NE tip of the southern 
arm of the inner spiral ($22^{\prime\prime}$ from the nucleus at 
${\rm PA}=64^{\circ}$).  Two additional peaks are seen $\sim45^{\prime\prime}$
(3~kpc) to the SW of the nucleus at ${\rm PA}=214^{\circ}$.  
The nonstellar emission map (Fig.~\ref{n1068ns}) shows that 
hot dust is present in these locations.  The subtraction of the nuclear PSF 
(right panel) suggests that there is little hot dust emission in the inner 
$\sim20^{\prime\prime}$ outside of the nucleus; widespread heating of dust 
to $\sim1000$~K can be ruled out.
It is curious that every peak of IR emission in the disk of NGC~1068, 
including the three such locations on the inner spiral as defined by 
the \citet{schinnerer} CO map, lie within the ionization cones of the AGN.

\subsection{Comparison to Other Wavelengths}

\begin{figure}
\vbox{
\begin{center}
\includegraphics[width=\textwidth]{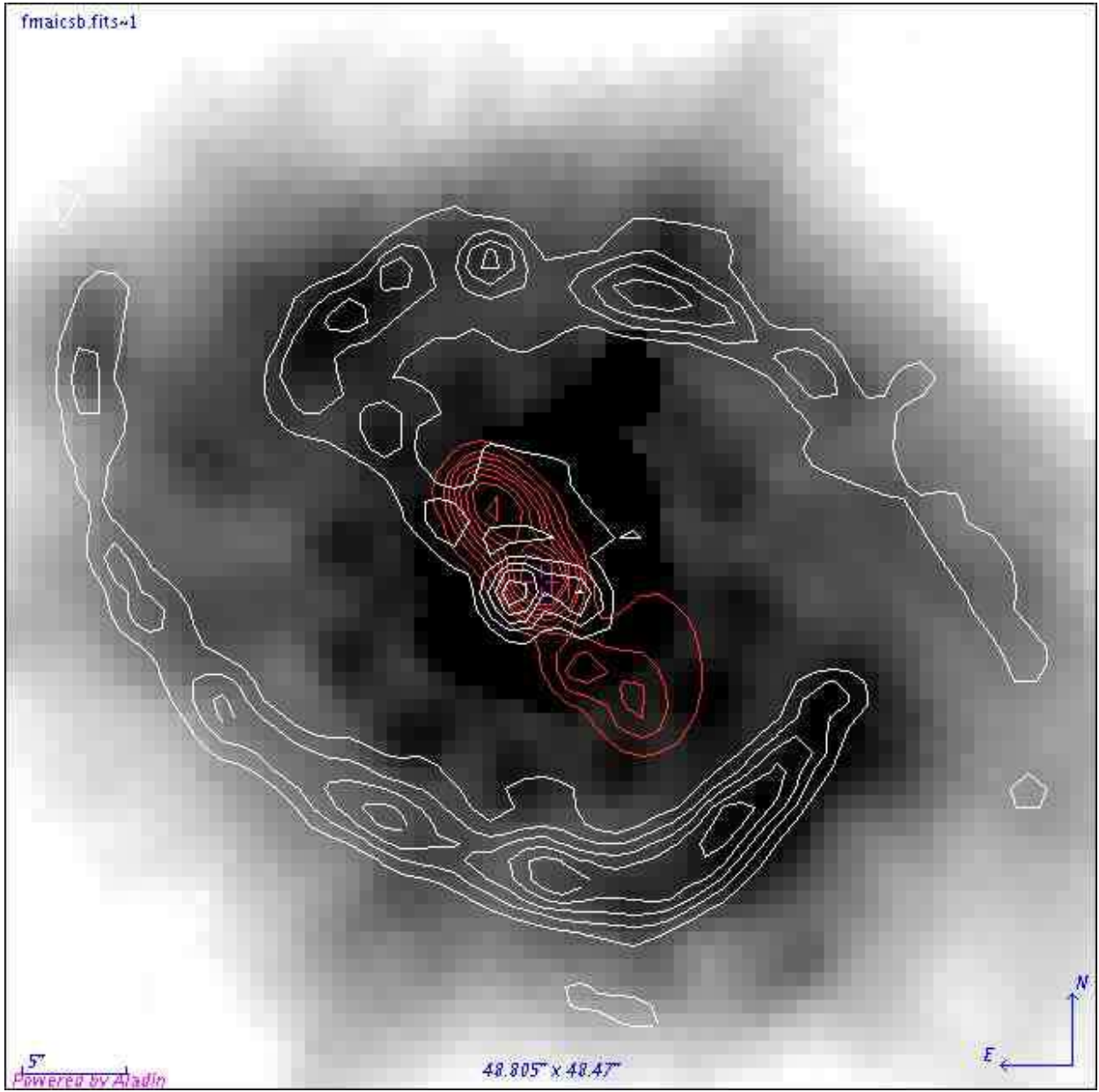}
\figcaption{\small
The $8~\mu$m image of the center of NGC~1068, with 1.49~GHz radio continuum 
contours from \citet{condon90} (red) and $^{12}{\rm CO}$~(1-0) contours 
from \citet{schinnerer} (white) overlaid.  Note that the peaks of $8~\mu$m 
emission lie at the locations where the radio jet axis intersects 
the inner spiral. 
\label{n1068radioco}
}
\end{center}}
\end{figure}

The radio continuum emission \citep{condon90} comes 
to an abrupt end at a distance of 500--700~pc from the nucleus, roughly 
halfway to the inner spiral, as shown in Fig.~\ref{n1068radioco}.  
The 8~$\mu$m knots on the inner spiral lie $\sim5^{\prime\prime}$ (300~pc)
farther along the radio axis.  \citet{cecil90} found that the high 
velocity nuclear outflow ends abruptly at the NE radio lobe and the SW 
radio hotspot ($4.5^{\prime\prime}$ from the nucleus).

The CO observations of \citet{schinnerer} (Fig.~\ref{n1068radioco}) 
show a coherent spiral structure.  The $8~\mu$m emission peaks 
coincide with regions of bright CO emission, though the converse 
is not true.  The two brightest PAH knots are seen at precisely the 
locations where the radio jet axis intersects the inner spiral 
(Fig.~\ref{n1068radioco}).  The 200--500~pc distance between the 
ends of the radio jets and the inner spiral indicate that the AGN 
is unlikely to trigger shock-induced star formation, however.

\begin{figure}
\vbox{
\begin{center}
\mbox{\includegraphics[width=3.5in]{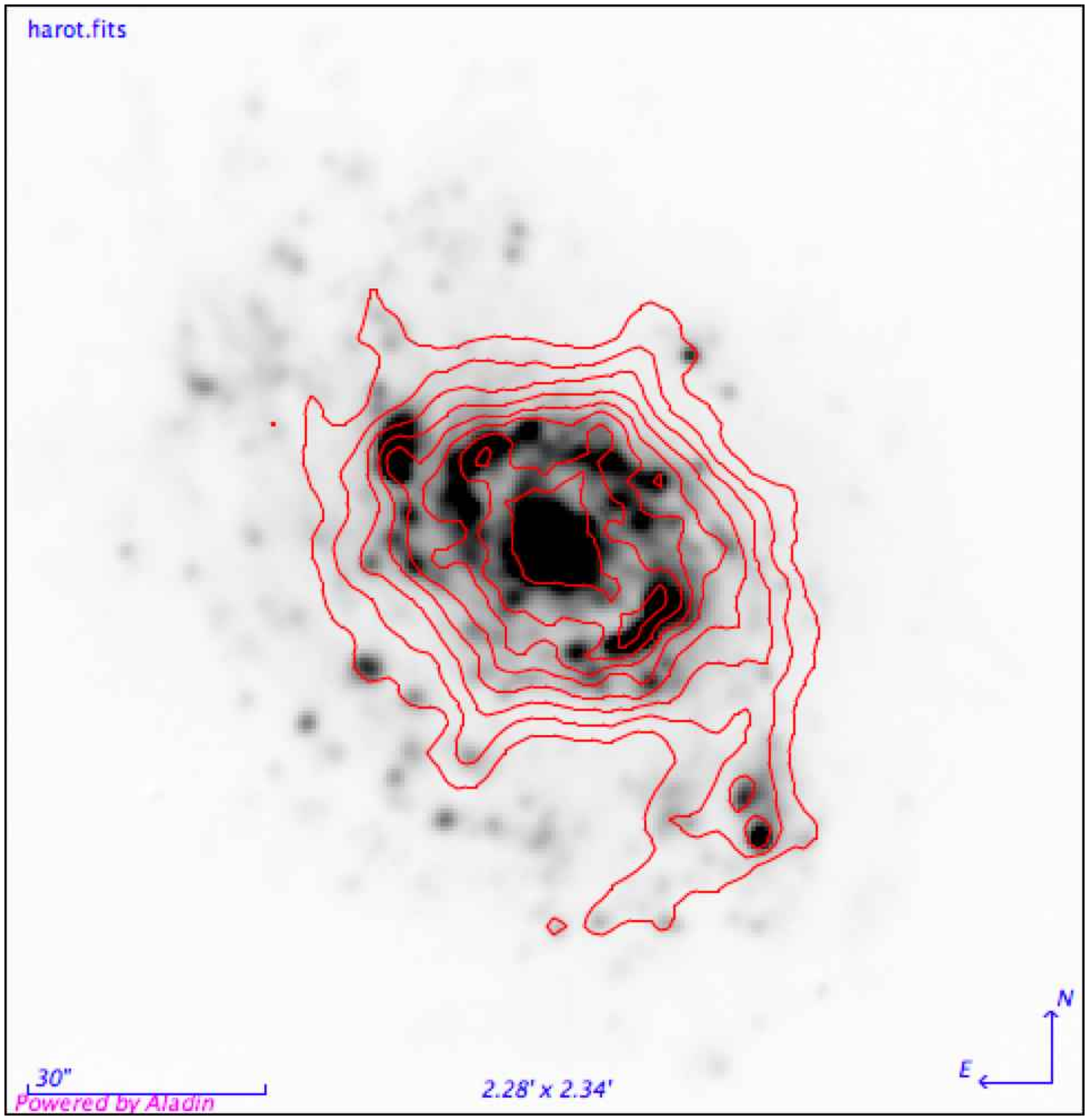}\quad
	\includegraphics[width=3.5in]{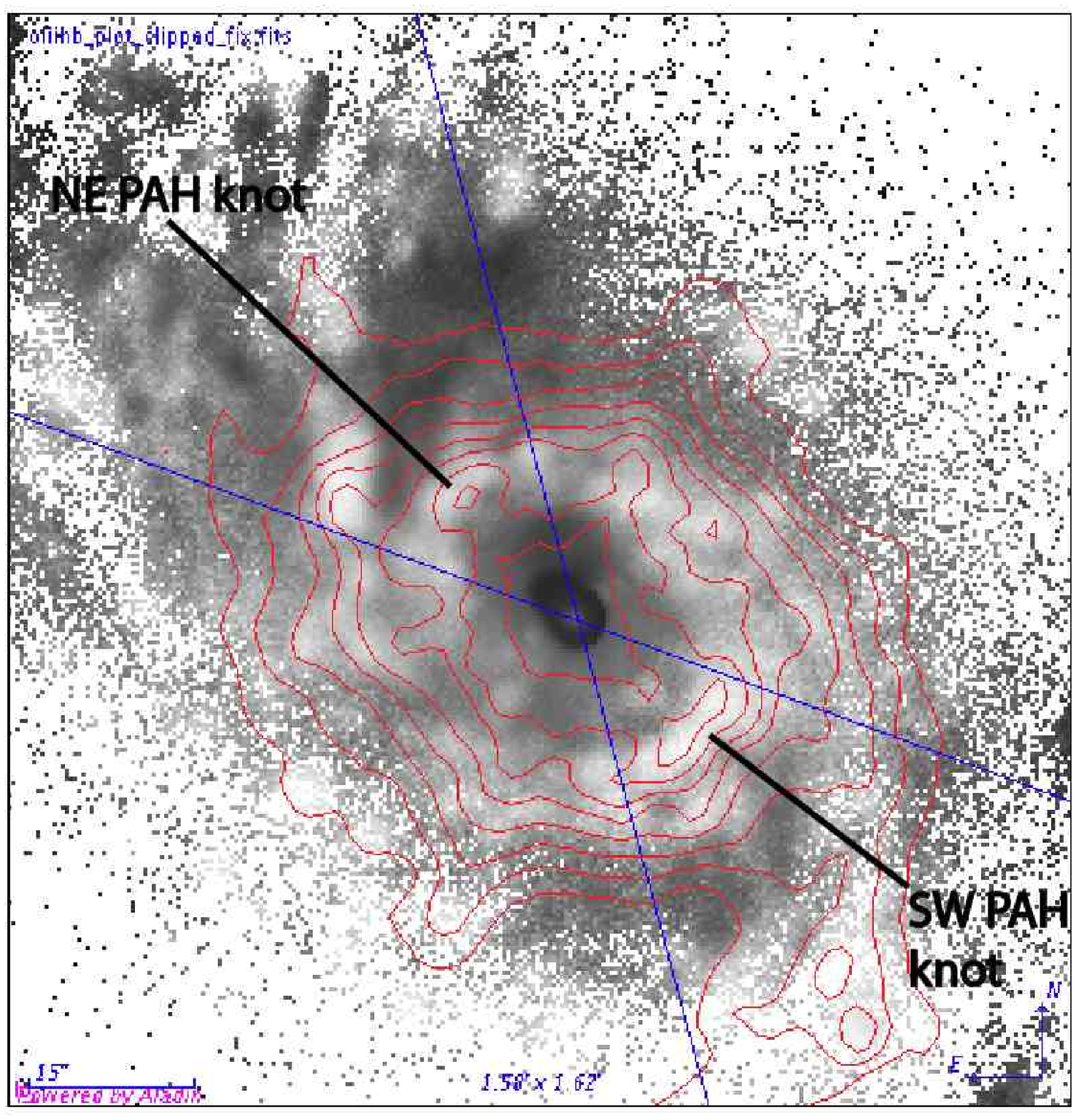}}
\figcaption{\small
Left: H$\alpha$ image of NGC~1068 from \citet{veilleux03} overlaid with 
$8~\mu$m contours.  Right:  [O~{\sc iii}]/H$\beta$ image of NGC~1068 from 
\citet{veilleux03} overlaid with $8~\mu$m contours.  The orientation of 
the ionization cones is shown, and the PAH knots on the inner spiral
are labeled.  The [O~{\sc iii}]/H$\beta$ ratio is 
low throughout the inner spiral (the light gray regions on the inner spiral
have ratios $\sim1$).  Elsewhere in the ionization cone, 
particularly to the NE, the [O~{\sc iii}]/H$\beta$ ratio is several times
as large (the dark gray regions have ratios $\sim7$), indicating that 
the AGN is the ionizing source.
\label{n1068line}
}
\end{center}}
\end{figure}

The optical emission line imaging of \citet{veilleux03} provides 
important insight into the origin of the observed PAH knots.  The PAH 
knots coincide with strong H$\alpha$ emission (Fig.~\ref{n1068line}, left 
panel) and with small [O~{\sc iii}]/H$\beta$ ratios (0.2--1.7; 
Fig.~\ref{n1068line}, right panel).  The $8~\mu$m and H$\alpha$ images show 
strikingly similar morphologies, tracing active star forming regions 
across a wide range in wavelength.  The [O~{\sc iii}]/H$\beta$ ratio is 
an indicator of mean ionization and temperature; high ratios ($>3$, as a 
rule of thumb) are typical of AGN, while lower ratios are typical of 
H~{\sc ii} regions \citep{osterbrock}.  As \citet{veilleux03} pointed 
out, high [O~{\sc iii}]/H$\beta$ ratios are seen out to $R\sim11$~kpc
in NGC~1068, most visibly in the NE ionization cone.  Figure~\ref{n1068line} 
clearly shows that both the starburst knots and the three fainter IR 
emission peaks have optical line ratios indicative of star formation.  
Although the 
H$\alpha$ and [O~{\sc iii}]/H$\beta$ data confirm the conclusion of 
\citet{lefloch01} that the PAH knots are due to star formation in those 
locations on the inner spiral, it is still surprising that bright PAH 
emission should be present from regions that are being illuminated 
by ionizing radiation from the AGN.

\begin{figure}
\vbox{
\begin{center}
\mbox{\includegraphics[width=3.5in]{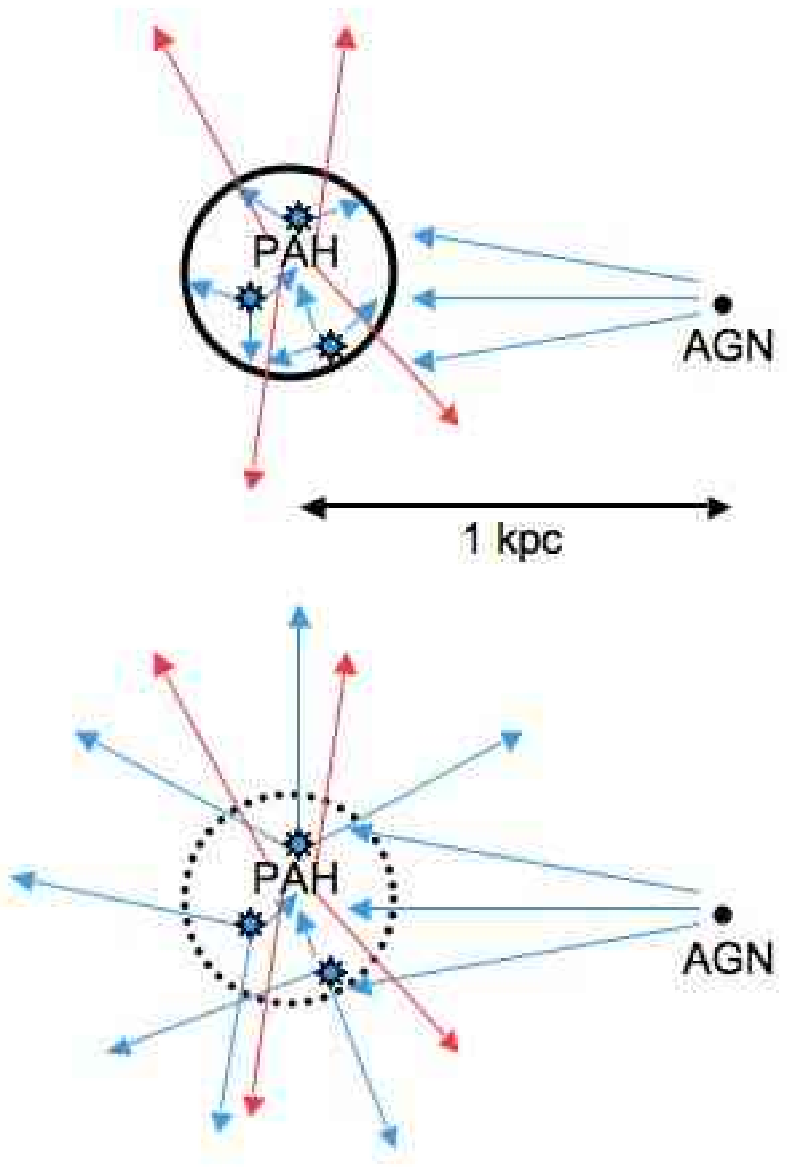}\quad
	\includegraphics[width=3.5in]{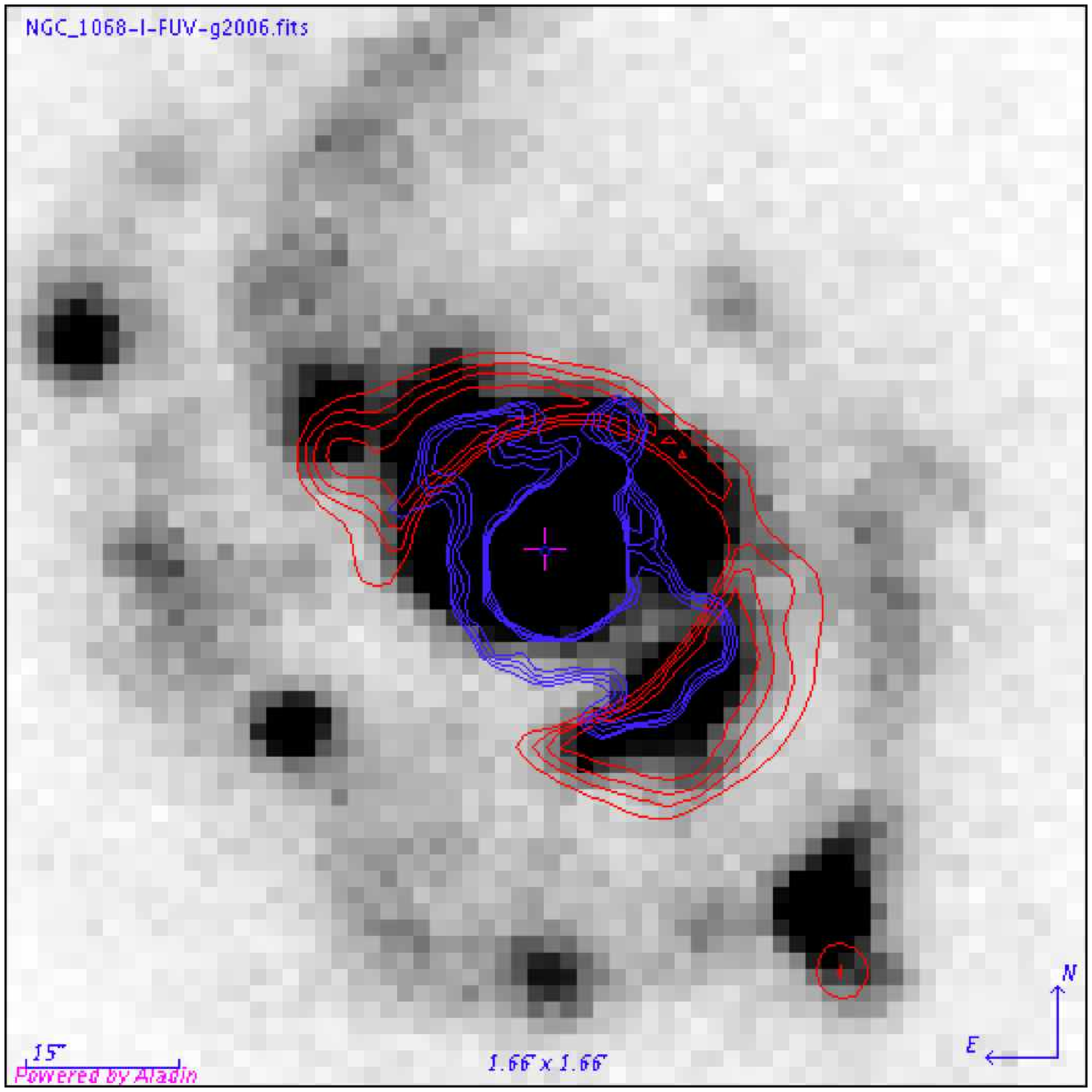}}
\figcaption{\small
Left: Diagram showing the optically thick (top) and optically thin (bottom)
scenarios described in the text.  The diagram is not to scale.  Blue lines 
represent UV photons from the AGN or massive stars; red lines represent 
PAH emission.
Right: The GALEX FUV (1516\AA; grayscale) data from \citet{galex} is 
compared with Spitzer $8\mu$m (blue) and PSF-subtracted $24\mu$m (red) 
data.  The regions of the inner spiral showing strong PAH and warm dust 
emission are also bright in the UV, indicating that these star forming 
regions are not optically thick to UV photons.  The slight offset 
($\sim3^{\prime\prime}$ NNE) between the GALEX and MIPS emission is real; 
the validity of the GALEX WCS solution was confirmed by plotting the 
positions of USNO catalog stars on the GALEX image.
\label{galexfig}
}
\end{center}}
\end{figure}

While the galaxy disk is illuminated by the AGN, it is unclear how opaque
the PAH emitting regions are to radiation from the AGN (e.g. see
Fig.~\ref{galexfig}, left panel).  The GALEX data \citep{galex}
provide little
guidance on this issue. Figure~\ref{galexfig} (right panel) is a GALEX 
FUV image of NGC~1068 shown
with contours of both the Spitzer 8$\mu$m and PSF-subtracted 24$\mu$m
emission overlaid.  As is apparent in the figure, the regions showing strong
PAH and warm dust emission are clearly UV luminous, indicating that (i) the
PAH emitting regions are optically thin to UV photons and thus unshielded by
the AGN, that (ii) the UV emanates from the optically thin outer envelope of
a much more embedded, optically thick starburst responsible for producing the
observed PAH emission, or (iii) that the UV emission is AGN light scattered
by optically thick clouds.  The latter possibility is unlikely --- a simple
assumption of a 100~pc spherical cloud with an albedo of one would reflect a
flux $\sim10^3$ fainter than the flux of the AGN.  Comparing measurements
within 6"(GALEX FWHM) radius apertures centered on the nucleus and on each
PAH knot,the GALEX data show that the PAH knots are $\sim 10$\% as bright as
the nucleus in the FUV and therefore scattering from the AGN is unlikely to
be a significant contribution. Thus, the UV emission is associated with
stars, but cannot be used to rule out whether the PAH regions are shielded
from the AGN.

Further interpretation 
is complicated by the near-coincidence between the axis of the AGN jet
and the position angle of the inner stellar bar.  Intense star formation 
is common at the ends of a bar \citep{combes85}, so the AGN may have no 
effect on the PAH knots at all.  It is also possible that the AGN is 
either triggering increased star formation or directly exciting PAH 
emission.  Either the high energy photons capable of destroying PAHs are 
absorbed near the AGN while lower energy photons are unobstructed 
over the entire $\sim11$~kpc of the illuminated disk, 
or the high energy photon flux within 1~kpc produces 
a PAH destruction rate less than or equal to the PAH formation rate in 
the star-forming regions.  For details on the parameters governing the 
formation and destruction of PAH molecules, see \citet{maloney99}.  
The radiative transfer model of \citet{siebenmorgen}
supports the latter hypothesis, finding that even in the optically thin 
regime PAHs survive exposure to a $10^{11}~{\rm L}_{\odot}$ AGN at distances 
of 100~pc or more.  However, \citet{voit92} calculated that PAHs could 
only survive direct exposure to a $2.5\times10^{10}~{\rm L}_{\odot}$ AGN 
at distances $\geq14$~kpc. 

\section{Conclusions}

We have investigated the influence of the AGN on PAH emission in 
NGC~1068, a LIRG with a Seyfert 2 nucleus.
Bright PAH knots are detected in the inner spiral in NGC~1068.  These 
knots lie directly along the axis of the ionization cone, which extends 
well beyond the inner spiral in the disk of the galaxy.  The knots also 
coincide with the ends of the inner stellar bar.  Contrary to 
the model of \citet{bv93}, the spectra and colors of the PAH knots are 
consistent with starlight and PAH emission, with very little if any hot dust.
Three other star forming regions are also detected: the NE tip of the 
southern arm of the inner spiral, and two regions 2.5--3~kpc to the SW.  
All three regions lie within the AGN ionization cones and show PAH emission 
at $8~\mu$m and a slight excess of $4.5~\mu$m emission relative to 
starlight, suggesting dust heated by young stars.  IRS spectra near the 
PAH knots indicate that PAHs are the dominant energy source, and show that 
PAH emission is present within a few hundred parsecs of the AGN.  PAH line 
ratios on the inner spiral are generally consistent with a cool, neutral 
ISM, while PAH line ratios interior to the inner spiral generally indicate 
a more ionized medium.  These data show that the AGN is not destroying all PAH 
molecules within the ionization cone.  The enhanced PAH emission within the 
ionization cones indicates that the AGN is either triggering increased star 
formation, directly exciting the PAHs, or has no effect on the dust 
properties in the ionization cones.

\acknowledgments

This research has made use of the NASA/IPAC Extragalactic Database (NED)
which is operated by the Jet Propulsion Laboratory, California Institute of
Technology, under contract with the National Aeronautics and Space
Administration.  We would like to thank J.~Condon for making his VLA 
radio continuum maps available to us, P.~Shopbell for H$\alpha$ and 
[O~{\sc iii}] images of NGC~1068, E.~Schinnerer for the CO image,
B.~Brandl for providing his average 
starburst galaxy spectrum, and J.~D.~Smith for assistance in using 
the PAHFIT software.  We thank the anonymous referee for constructive 
comments.
Support for this work was provided by NASA
through contracts 1263752, 1264790, and 1267948 (D.C.K and S.V.)
issued by JPL/Caltech.


\clearpage

\begin{deluxetable}{lllllllll}
\tabletypesize{\scriptsize}
\tablecolumns{9}
\tablewidth{0pt}
\tablecaption{Spitzer Flux Density Measurements}
\tablehead{
\multicolumn{1}{l}{} &
\multicolumn{4}{c}{IRAC}   &
\multicolumn{1}{c}{} &  
\multicolumn{2}{c}{MIPS} & \\
\cline{3-6} \cline{8-9}\\
\colhead{Feature} & \colhead{Aperture Radius} & \colhead{$3.6~\mu$m} & \colhead{$4.5~\mu$m} & \colhead{$5.8~\mu$m} & \colhead{$8~\mu$m} & & \colhead{$24~\mu$m} & \colhead{$70~\mu$m} \\
\colhead{} & \colhead{arcsec (kpc)} & \colhead{Jy} & \colhead{Jy} & \colhead{Jy} & \colhead{Jy} & & \colhead{Jy} & \colhead{Jy} \\
}
\startdata
NGC 1068: Total & $90^{\prime\prime}$ (6 kpc) & 3.8 & 5.1 & 13 & 23 & & 80 & 180 \\
NGC 1068: 1 kpc & $15^{\prime\prime}$ & 2.8 & 4.3 & \phantom{0}9.0 & 17 & & ... & ... \\
NGC 1068: NE PAH knot & $4^{\prime\prime}$ (0.26 kpc)$^{\rm a}$ & 0.011 & 0.006 & \phantom{0}0.05 & \phantom{0}0.11 & & ... & ... \\
NGC 1068: SW PAH knot & $5^{\prime\prime}$ (0.33 kpc)$^{\rm a}$ & 0.024 & 0.020 & \phantom{0}0.11 & \phantom{0}0.26 & & ... & ... \\
NGC 1068: Inner Spiral & $12^{\prime\prime}$--$18^{\prime\prime}$ (0.8--1.2 kpc) & 0.29 & 0.20 & \phantom{0}0.80 & \phantom{0}1.85 & & ... & ... \\
\\
\enddata
\tablenotetext{a}{Effective radius of custom aperture.  Flux densities for 
the PAH knots are relative to the local background elsewhere in the inner
spiral.}
\tablecomments{The approximate uncertainties are indicated by the number of 
significant figures; specifically, 10\% at IRAC wavelengths, and 20\% at $70~\mu$m.  See text for a discussion of the $24~\mu$m uncertainties.}
\end{deluxetable}

\begin{deluxetable}{lllllllllll}
\tabletypesize{\tiny}
\tablecolumns{11}
\tablewidth{0pt}
\tablecaption{Emission Features (Low Resolution)}
\tablehead{
\colhead{Aperture} & \colhead {r (kpc)} & \colhead{PAH 6.2} & \colhead{[Ar~II] 6.98} & \colhead{PAH 7.7} & \colhead{PAH 8.6} & \colhead{${\rm H}_2$~9.66} & \colhead{[S~IV] 10.51} & \colhead{PAH 11.3} & \colhead{PAH 12.7} & \colhead{[Ne~II] 12.8} \\ 
\colhead{($\Delta {\rm RA}, \Delta {\rm DEC}$)} & & & & & & & & & & \\
}
\startdata
\multicolumn{11}{c}{Inner Spiral}\\
1 & 0.95 & $328\pm6$ & $12\pm2$ & $1020\pm20$ & $175\pm3$ & $9.0\pm0.8$ & $5.1\pm0.6$ & $285\pm5$ & $170\pm2$ & $35.7\pm0.4$ \\
($4.3^{\prime\prime}, 13.5^{\prime\prime}$) & & 2.04 & 0.042 & 3.91 & 0.65 & 0.066 & 0.032 & 1.08 & 0.60 & 0.19 \\
2 & 0.88 & $250\pm4$ & $9\pm2$ & $780\pm10$ & $118\pm3$ & $7\pm1$ & $6.8\pm0.5$ & $228\pm2$ & $124\pm2$ & $22.0\pm0.4$ \\
($0.4^{\prime\prime}, 13.2^{\prime\prime}$) & & 1.71 & 0.066 & 3.14 & 0.43 & 0.043 & 0.039 & 0.80 & 0.42 & 0.11 \\
3 & 0.87 & $268\pm8$ & $8\pm3$ & $820\pm30$ & $130\pm10$ & $11\pm3$ & $5\pm2$ & $226\pm8$ & $94\pm8$ & $22.9\pm0.9$ \\
($-5.4^{\prime\prime}, 11.9^{\prime\prime}$) & & 0.55 & 0.024 & 1.73 & 0.32 & 0.026 & 0.029 & 0.75 & 0.29 & 0.11 \\
4 & 0.90 & $236\pm4$ & $9\pm2$ & $770\pm10$ & $127\pm4$ & $10\pm1$ & $3.9\pm0.6$ & $245\pm3$ & $112\pm2$ & $24.2\pm0.5$ \\
($-10.1^{\prime\prime}, 9.0^{\prime\prime}$) & & 1.22 & 0.058 & 2.85 & 0.49 & 0.088 & 0.028 & 0.96 & 0.36 & 0.12 \\
5 & 0.98 & $221\pm4$ & $6\pm3$ & $560\pm20$ & $153\pm4$ & $19\pm2$ & $5\pm1$ & $275\pm5$ & $100\pm2$ & $20.8\pm0.4$ \\
($-14.0^{\prime\prime}, 4.5^{\prime\prime}$) & & 1.03 & 0.031 & 1.69 & 0.57 & 0.21 & 0.042 & 1.20 & 0.31 & 0.10 \\
9 & 0.98 & $268\pm3$ & $13\pm2$ & $950\pm10$ & $143\pm3$ & $4.5\pm0.8$ & $7.7\pm0.6$ & $287\pm2$ & $134\pm2$ & $28.5\pm0.4$ \\
($-2.5^{\prime\prime}, -14.4^{\prime\prime}$) & & 2.04 & 0.089 & 3.50 & 0.48 & 0.030 & 0.043 & 0.92 & 0.39 & 0.12 \\
10 & 1.02 & $125\pm3$ & ... & $644\pm9$ & $82\pm2$ & ... & $4.2\pm0.3$ & $143\pm1$ & $55\pm1$ & $10.7\pm0.3$ \\
($4.3^{\prime\prime}, -14.7^{\prime\prime}$) & & 0.93 & ... & 4.84 & 0.47 & ... & 0.021 & 0.42 & 0.13 & 0.038 \\
11 & 1.02 & $148\pm3$ & $5\pm1$ & $570\pm10$ & $79\pm2$ & $2.9\pm0.6$ & $5.6\pm0.4$ & $199\pm1$ & $127\pm1$ & $13.4\pm0.3$ \\
($14.8^{\prime\prime}, -3.8^{\prime\prime}$) & & 0.67 & 0.032 & 2.18 & 0.30 & 0.021 & 0.035 & 0.78 & 0.48 & 0.075 \\
\hline 
\multicolumn{11}{c}{Inside the Inner Spiral}\\
6 & 0.68 & $133\pm3$ & $6\pm2$ & $630\pm20$ & $105\pm3$ & $2.0\pm0.5$ & $4.2\pm0.4$ & $155\pm2$ & $86\pm2$ & $15.0\pm0.4$ \\
($-10.1^{\prime\prime}, -1.8^{\prime\prime}$) & & 0.24 & 0.018 & 1.29 & 0.25 & 0.008 & 0.019 & 0.50 & 0.30 & 0.079 \\
7 & 0.65 & $179\pm6$ & $4\pm2$ & $830\pm20$ & $133\pm4$ & $8.1\pm0.8$ & $12.0\pm0.5$ & $244\pm3$ & $100\pm3$ & $23.7\pm0.4$ \\
($-7.2^{\prime\prime}, -6.5^{\prime\prime}$) & & 0.22 & 0.008 & 1.12 & 0.22 & 0.030 & 0.039 & 0.46 & 0.15 & 0.053 \\
8 & 0.74 & $126\pm4$ & $1.7\pm0.9$ & $660\pm20$ & $137\pm3$ & $4.1\pm0.5$ & $7.2\pm0.4$ & $223\pm2$ & $117\pm2$ & $19.1\pm0.3$ \\
($-5.4^{\prime\prime}, -9.7^{\prime\prime}$) & & 0.20 & 0.004 & 1.20 & 0.29 & 0.016 & 0.029 & 0.60 & 0.29 & 0.069 \\
13 & 0.68 & $157\pm4$ & $5\pm1$ & $670\pm20$ & $120\pm3$ & $4.3\pm0.4$ & $7.2\pm0.4$ & $171\pm2$ & $101\pm2$ & $17.1\pm0.3$ \\
($2.2^{\prime\prime}, 10.0^{\prime\prime}$) & & 0.37 & 0.017 & 1.56 & 0.30 & 0.018 & 0.031 & 0.48 & 0.26 & 0.064 \\
14 & 0.66 & $146\pm5$ & $5\pm2$ & $640\pm20$ & $103\pm3$ & $2.8\pm0.4$ & $6.3\pm0.4$ & $150\pm2$ & $91\pm3$ & $15.5\pm0.4$ \\
($-2.2^{\prime\prime}, 9.7^{\prime\prime}$) & & 0.32 & 0.019 & 1.56 & 0.26 & 0.011 & 0.024 & 0.39 & 0.25 & 0.063 \\
15 & 0.64 & $141\pm5$ & $5\pm2$ & $620\pm20$ & $104\pm4$ & $3.3\pm0.5$ & $5.4\pm0.4$ & $125\pm2$ & $71\pm3$ & $16\pm6$ \\
($-6.8^{\prime\prime}, 6.8^{\prime\prime}$) & & 0.30 & 0.016 & 1.47 & 0.26 & 0.012 & 0.021 & 0.33 & 0.19 & 0.066 \\
17 & 0.78 & $95\pm3$ & $1.8\pm0.7$ & $480\pm10$ & $83\pm2$ & $1.6\pm0.4$ & $4.2\pm0.3$ & $116\pm1$ & $54\pm2$ & $10.6\pm0.3$ \\
($11.5^{\prime\prime}, -1.6^{\prime\prime}$) & & 0.23 & 0.006 & 1.19 & 0.22 & 0.007 & 0.019 & 0.39 & 0.19 & 0.056 \\ 
\hline 
\multicolumn{11}{c}{Apertures Nearest the Nucleus}\\
16 & 0.54 & $99\pm5$ & $2\pm1$ & $630\pm20$ & $41\pm4$ & $7.4\pm0.3$ & $28.6\pm0.7$ & $187\pm2$ & $48\pm3$ & $18.7\pm0.7$ \\
($-2.2^{\prime\prime}, -7.8^{\prime\prime}$) & & 0.11 & 0.003 & 0.64 & 0.04 & 0.010 & 0.031 & 0.11 & 0.02 & 0.012 \\
18 & 0.48 & $165\pm4$ & $12\pm1$ & $690\pm10$ & $52\pm3$ & $2.5\pm0.6$ & $32.5\pm0.4$ & $296\pm2$ & $306\pm3$ & $34.0\pm0.8$ \\
($6.1^{\prime\prime}, 3.7^{\prime\prime}$) & & 0.24 & 0.024 & 0.84 & 0.05 & 0.003 & 0.030 & 0.16 & 0.14 & 0.023 \\
19 & 0.54 & $119\pm4$ & $6\pm2$ & $600\pm20$ & $88\pm4$ & $2.7\pm0.5$ & $5.3\pm0.5$ & $133\pm2$ & $92\pm2$ & $11.8\pm0.6$ \\
($-7.9^{\prime\prime}, 1.7^{\prime\prime}$) & & 0.18 & 0.015 & 0.98 & 0.18 & 0.011 & 0.022 & 0.34 & 0.19 & 0.035 \\
\hline 
\multicolumn{11}{c}{Outside the Inner Spiral}\\
12 & 1.37 & $35\pm2$ & ... & $173\pm7$ & $26\pm2$ & $0.7\pm0.3$ & ... & $45.8\pm0.8$ & $16\pm1$ & $2.3\pm0.2$ \\
($17.3^{\prime\prime}, -11.1^{\prime\prime}$) & & 0.90 & ... & 2.22 & 0.26 & 0.010 & ... & 0.42 & 0.15 & 0.031 \\
\enddata
\tablecomments{Fluxes are in units of $10^{-21}~{\rm W~cm^{-2}}$, with 
the equivalent width in microns below.}
\end{deluxetable}

\begin{deluxetable}{llc}
\tabletypesize{\scriptsize}
\tablecolumns{3}
\tablewidth{0pt}
\tablecaption{Nuclear Emission Features (High Resolution)}
\tablehead{
\colhead{Line} & \colhead {Flux} & \colhead{Equivalent width} \\
\colhead{} & \colhead{$10^{-21}~{\rm W~cm}^{-2}$} & \colhead{$\mu$m} \\
}
\startdata
[S~IV] 10.51 & $\phantom{0}460\pm60$ & 0.006 \\
${\rm [Ne~II]}~12.81$ & $\phantom{0}520\pm60$ & 0.008 \\
${\rm [Ne~V]}~14.3$ & $1110\pm60$ & 0.017 \\
${\rm [Ne~III]}~15.55$ & $2100\pm200$ & 0.024 \\
${\rm [S~III]}~18.71$ & $\phantom{0}220\pm20$ & 0.004 \\
${\rm [Ne~V]}~24.31$ & $\phantom{0}800\pm100$ & 0.023 \\
${\rm [O~IV]}~25.89$ & $2200\pm200$ & 0.077 \\
${\rm [S~III]}~33.48$ & $\phantom{0}420\pm60$ & 0.025 \\
${\rm [Si~II]}~34.81$ & $\phantom{0}620\pm20$ & 0.034 \\
${\rm [Ne~III]}~36.0$ & $\phantom{0}130\pm20$ & 0.006 \\
\enddata
\end{deluxetable}

\clearpage

\end{document}